\documentclass{ws-rmp}
\usepackage{latexsym}
\usepackage{amsmath}
\usepackage{amssymb}
\usepackage{amscd}
\usepackage{amsfonts}
\usepackage{amstext}
\usepackage{graphicx}
\usepackage[all]{xy}
\usepackage{float}
\usepackage{color}

\newtheorem{fact}{Fact}
\usepackage{stmaryrd}

\begin{document}
\global\long\global\long\global\long\def\bra#1{\mbox{\ensuremath{\langle#1|}}}
\global\long\global\long\global\long\def\ket#1{\mbox{\ensuremath{|#1\rangle}}}
\global\long\global\long\global\long\def\bk#1#2{\mbox{\ensuremath{\ensuremath{\langle#1|#2\rangle}}}}
\global\long\global\long\global\long\def\kb#1#2{\mbox{\ensuremath{\ensuremath{\ensuremath{|#1\rangle\!\langle#2|}}}}}

\title{Convexity of momentum map, Morse index, and quantum entanglement}



\author{ADAM SAWICKI$^{1,2}$, MICHA{\L} OSZMANIEC\footnote{oszmaniec@cft.edu.pl}\ \ $^{2}$\,  MAREK KU\'S$^{2}$}

\address{$^1$School of Mathematics, University of Bristol, \\
University Walk, Bristol BS8 1TW, UK \\
$^2$Center for Theoretical Physics, Polish Academy of Sciences, \\
Al. Lotnik\'ow 32/46, 02-668 Warszawa, Poland}

\maketitle

\begin{abstract}
We analyze form the topological perspective the space of all SLOCC
(Stochastic Local Operations with Classical Communication)
classes of pure states for composite quantum systems. We do it for both
distinguishable and indistinguishable particles. In general, the topology of
this space is rather complicated as it is a non-Hausdorff space. Using
geometric invariant theory (GIT) and momentum map geometry we propose a way
to divide the space of all SLOCC classes into mathematically and physically
meaningful families. Each family consists of possibly many `asymptotically'
equivalent SLOCC classes. Moreover, each contains exactly one distinguished
SLOCC class on which the total variance (a well defined
measure of entanglement) of the state $\mathrm{Var}[v]$ attains maximum. We
provide an algorithm for finding critical sets of  $\mathrm{Var}[v]$, which
makes use of the convexity of the momentum map and allows classification of
such defined families of SLOCC classes. The number of families is in general
infinite. We introduce an additional refinement into finitely many groups of
families using the recent developments in the momentum map geometry known as
Ness stratification. We also discuss how to define it equivalently using the
convexity of the momentum map applied to SLOCC classes. Moreover, we note
that the Morse index at the critical set of the total variance of state has
an interpretation of number of non-SLOCC directions in which entanglement
increases and calculate it for several exemplary systems. Finally, we
introduce the SLOCC-invariant measure of entanglement as a square root of the
total variance of state at the critical point and explain its geometric
meaning.
\end{abstract}

\maketitle

\section{Introduction}

The problem of classification of pure multipartite entanglement is a
recurrent subject in quantum information theory
\cite{ChuangNielsen,Horodeccy}. There are two possible ways to approach it.
The idea behind the first one is based on a perhaps trivial but very fruitful
observation that entanglement has to be invariant under local unitary
operations, i.e.\ unitary operations applied independently to every subsystem
of a given system. In terms of group theory this amounts to the study of
orbits of the compact group $K=SU(N_{1})\times\ldots\times SU(N_{L})$ on the
compact manifold $M=\mathbb{P}(\mathcal{H})$, where $\mathcal{H}=
\mathcal{H}_{1}\otimes\cdots\otimes\mathcal{H}_{L}$ is the Hilbert space of
the whole system and $\mathbb{P}(\mathcal{H})$
is the corresponding complex projective space %
\footnote{For indistinguishable particles we need a slight modification of this
setting as it will be explained in Section~\ref{sec:indistinguishable}%
}. Since all such orbits are closed the quotient space $M/K$ is at least a
topological Hausdorff\footnote{A topological space is Hausdorff if any two
points can be separated by open sets.} space and one can perform the desired
classification by searching for $K$-invariant polynomials
\cite{Sudbery,Vrana}. Although the structure of invariant polynomials in well
known\footnote{Invariant polynomials are given by expressions of the form
$\mathrm{tr}\left(\kb{\psi^{\otimes k}}{\psi^{\otimes k}}X\right),$ where $X$
commutes with the diagonal action of $K$ on $\mathcal{H}^{\otimes k}$.}, it
is in general a difficult problem to find the minimal generating set of this
ring. Another disadvantage of this method is that the rich geometry of the
problem is barely visible in this approach. It is well known that the complex
projective space is a K\"ahler manifold and hence a symplectic manifold with
the Fubini-Study symplectic form $\omega_{FS}$. Although $M$ is symplectic
the restriction of $\omega_{FS}$ to an orbit of $K$ is typically not
symplectic. Inspired by this observation we have recently introduced a
discrete entanglement measure defined as the degree of degeneracy of
$\omega_{FS}$ restricted to $K$-orbits \cite{sawicki11}. In simple words we
showed that more non-symplectic $K$-orbit through the state $[v]$ is, more
entangled is the state $[v]$. The key ingredient in the proof of this
statement was the concept of the momentum map
$\mu:\mathbb{P}(\mathcal{H})\rightarrow\mathfrak{k}^{\ast}$, where
$\mathfrak{t}^{\ast}$ is the dual space to the Lie algebra $\mathfrak{k}$ of
the group $K$. In our setting this map encodes the information about
expectation values of the local (one-particle) observables, i.e.\ it is given
by the collection of reduced one-particle density matrices. The degree of the
degeneracy of $\omega_{FS}$ at $[v]$ has thus a physical interpretation - is
the dimension of the set of $K$-equivalent states which have the same reduced
one-particle density matrices. This degeneracy can be thus interpreted as a
`measure' of an additional information which one has to gain in order to
distinguish state $[v]$ from the states which are $K$-equivalent to $[v]$
using only one-particle density matrices.

The second approach to quantum entanglement uses the paradigm that two pure
quantum states can in principle be used for the same quantum computations if
and only if they can be transformed into each other by the class of so-called
SLOCC operations \cite{Vidal}. Invertible SLOCC operations are represented by
$G=SL(N_{1},\mathbb{C})\times\ldots\times SL(N_{L},\mathbb{C})$ and
$G=K^{\mathbb{C}}$, i.e.\ the group $G$ is the complexification of the group
$K$. In the language of the group theory this means that $G$ is a reductive
group. However, since $G$ is not compact its orbits in $M$ may not be closed
and hence the space $M/G$ is typically non-Hausdorff. Nevertheless, the
celebrated theorem of Hilbert and Nagata \cite{M03} ensures that the ring of
invariant polynomials is finitely generated and it is possible to distinguish
between different \textbf{closed} $G$-orbits in $M$.

In the current paper we focus on the investigation of SLOCC classes of pure
states. As the problem of finding explicitly all SLOCC classes is regarded
intractable we propose division of the set of all $G$-orbits into families.
It stems from the structure of the $G$-orbits space which is a non-Hausdorff
space. We make an extensive use of the geometric invariant theory and
momentum map geometry which we review in sections \ref{sec:momentum-map} and
\ref{sec:stratification}. The essence of our approach is the following. We
look at the total variance of the state, $\mathrm{Var}[v]$, which was
recently introduced by Klyachko as a measure of entanglement
\cite{klyachko07}. The total variance takes the maximal value for maximally
entangled states and the minimal value for separable states. Using the
Riemannian metric on $\mathbb{P}(\mathcal{H})$ we construct the gradient
vector field of $\mathrm{Var}[v]$. Moving the state along the gradient flow
always increases the total variance. The limit sets of this flow generated by
this field are exactly critical sets of $\mathrm{Var}[v]$. Making use of the
momentum map geometry we notice that the function $\mathrm{Var}[v]$ is up to
an additive constant the minus square norm of the momentum map $||\mu||^{2}$.
The critical points of the later were studied in early 1980's by Kirwan and
Ness \cite{Kirwan82,Ness84}. Translating their results into language of
entanglement we find that each critical set of $\mathrm{Var}[v]$ can be
characterized as a unique $K$-orbit inside $G$-orbit on which
$\mathrm{Var}[v]$ attains maximum.  Moreover, the gradient of
$\mathrm{Var}[v]$ is tangent to $G$-orbits, and hence its flow can be
realized by local operations. The division of SLOCC classes into families is
the following. We identify all states which are carried by the gradient flow
into one critical $K$-orbit. These might be not only all states from the
$G$-orbit that contains the chosen critical $K$-orbit, but also those which
can be taken to it in the limit. Nevertheless, each so-defined family of
SLOCC classes can be represented by exactly one SLOCC class passing through
the critical set of the total variance. Taking into account the physical
meaning of $\mathrm{Var}[v]$ it seems to be a natural procedure.

In order to determine all above described families we need an effective
algorithm for finding critical sets of the total variance. As we explain in
section \ref{sec:The-SLOCC-invariant} this problem can be divided into two
essentially different parts. The first one boils down to finding all states
for which reduced single-particle density matrices are maximally mixed. The
second problem concerns finding critical sets of $\mathrm{Var}[v]$ among
states whose closures of  $G$-orbits in the Hilbert space $\mathcal{H}$
contain the zero vector. We show that this task can be greatly simplified by
using the convexity property of the momentum map. Essentially, it reduces to
considerations of eigenspaces of a matrix constructed from the
single-particle reduced density matrices.

In general the number of so-defined SLOCC families is infinite. Using Ness
stratification, which we discuss in section \ref{sub:Startification}, we
propose a way to divide them into finitely many groups. Each group consists
of families represented by critical sets of $\mathrm{Var}[v]$ with the same
spectra of reduced one-particle density matrices. Next, we show that the idea
of the momentum map can be useful for equivalent characterization of these
groups. The ordered spectra of reduced one-particle density matrices
corresponding to the states from the closure of $G$-orbits form a convex
polytope, so-called SLOCC Kirwan polytope . Moreover, the number of such
polytopes is finite, even if the number of $G$-orbits is infinite. As we
describe in section \ref{sec:stratification} it leads to an equivalent
definition for the group of families of SLOCC classes as all states whose
SLOCC Kirwan polytopes have the same closest point to the origin. We also
notice that the value of $\mathrm{Var}[v]$, or actually square root of it,
calculated in the critical point can be understood geometrically in terms of
distance of the SLOCC Kirwan polytope from the origin. This introduces a
hierarchy among different groups of families of SLOCC classes. Finally, we
point out that the Morse index at a critical set have a nice physical
interpretation of number of non-SLOCC directions in which entanglement
increases.

We illustrate our considerations by calculating in sections
\ref{sec:The-main-examples} and \ref{sec:indistinguishable}
several examples for both distinguishable and indistinguishable particles. We
point out that in some cases (bipartite, three qubits) all SLOCC classes contain critical set of the total variance. For
these cases our classification and the one known in literature overlap.
However, for four qubits the situation is very different. We do have families
of SLOCC classes which contain more than one $G$-orbit and in section 5.3 we
show which are of these kind. We also prove that even in this situation there
are still families which consists of a single $G$-orbit. We point out that
this phenomenon is connected to stable \emph{vs} semistable classification of
points in GIT. In section \ref{two-state bosons} we also show that for system
of arbitrary many two-state bosons all groups of families of SLOCC classes,
except the one given by maximally mixed reduced one-particle density
matrices, contain exactly one critical $K$-orbit. In the course of
argumentation we exhibit a connection between the norm square of the momentum
map $\|\mu\|^{2}$, the total variance of state $\mathrm{Var}[v]$, and the
second order Casimir invariant for the group $K$. The later allows for an
identification of them as an expectation value of an observable. That makes
them, at least in principle, experimentally measurable.

\section{Geometry of Momentum map on the complex projective space}
\label{sec:momentum-map}

In this part we briefly discuss the concept of the momentum map. We present
the general definition and the construction of the momentum map on the
complex projective space. Finally we discuss in more detail convexity
properties of the momentum map. For an extensive review we refer to
\cite{GS84}.

Let $K$ be a compact connected semisimple Lie group. Let us assume that $K$
acts on a symplectic manifold $(M,\omega)$ in the symplectic way, i.e.\
leaving the symplectic form $\omega$ invariant,
\begin{gather*}
\Phi:K\times M\rightarrow M,\,\,\Phi_{g}=\Phi(g,\cdot),\,\,\,
\Phi_{g}^{\ast}\omega=\omega.
\end{gather*}
For such an action there exists a momentum map \cite{GS90}, i.e.\ the unique
map, $\mu:M\rightarrow\mathfrak{k}^{\ast}$, from the manifold $M$ to the dual
space $\mathfrak{k}^{\ast}$ of the Lie algebra $\mathfrak{k}=Lie(K)$, such
that
\begin{enumerate}
\item The map $\mu$ is equivariant, i.e.\
    $\mu(\Phi_{g}(x))=\mathrm{Ad}_{g}^{\ast}\mu(x)$, where
    $\mathrm{Ad}_{g}^{\ast}$
is coadjoint action
$\langle\mathrm{Ad}_{g}^{\ast}\alpha,\,\xi\rangle=\langle\alpha,\mathrm{Ad}_{g^{-1}}\xi\rangle=\langle\alpha,\,
g^{-1}\xi g\rangle$, where $\langle\,\cdot,\cdot\rangle$ is the pairing
between $\mathfrak{k}$ and $\mathfrak{k}^{\ast}$, and $\mathrm{Ad}_{g}$
denotes the adjoint action of $K$ on $\mathfrak{k}$, i.e.\
 $\mathrm{Ad}_{g}\xi=g\xi g^{-1}$ for $g\in K$ and $\xi\in\mathfrak{k}$.
\item For any $\xi\in\mathfrak{k}$, the fundamental vector field,
\begin{equation}
\hat{\xi}(x)=\frac{d}{dt}\Big|_{t=0}\Phi_{exp(t\xi)}(x)\label{fundvectf}
\end{equation}
is the Hamiltonian vector field of the Hamiltonian function $\mu_{\xi}(x)
=\langle\mu(x),\,\xi\rangle$, i.e.\ $d\mu_{\xi}=\omega(\hat{\xi},\cdot)$.
\end{enumerate}

\subsection{Momentum map on the complex projective space}

For our purposes we need the momentum map on the complex projective space,
$\mathbb{P}(\mathcal{H})$. Before presenting its construction let us review
symplectic properties of $\mathbb{P}(\mathcal{H})$. Let $G$ be a complex
reductive group, i.e.\ it is the complexification, $G=K^{\mathbb{C}}$, of its
maximal compact subgroup $K$. Let us assume that $G$ acts linearly on
$\mathcal{H}\simeq\mathbb{C}^{N}$. For simplicity of presentation we will
assume that $G$ is actually a subgroup of $SL(\mathcal{H},\mathbb{C})$ and
hence $K$ can be chosen as a subgroup of $SU(\mathcal{H})$. Denote by
$\bk{\cdot}{\cdot}$ the $K$-invariant scalar product on $\mathcal{H}.$ The
action of $G$ on $\mathcal{H}$ induces in a natural way an action of $G$, and
hence of $K$, on the complex projective space $\mathbb{P}(\mathcal{H})$,
$g[v]=[gv]$. In the notation we took advantage of the linearity of the group
action on $\mathcal{H}$ and we will consequently use $gv$ instead of
$\Phi_{g}(v)$ in this case.

For further reference let us make two remarks.
\begin{enumerate}
\item Since the group $SU(\mathcal{H})$ acts transitively on
    $\mathbb{P}(\mathcal{H})$ the tangent space
    $T_{[v]}\mathbb{P}(\mathcal{H})$ at $[v]$ is spanned by the
    fundamental vector fields (\ref{fundvectf}) where
    $\Phi_{exp(t\xi)}[v]=[e^{t\xi}v]$,
    $\xi\in\mathfrak{su}(\mathcal{H})=Lie(SU(\mathcal{H}))$ (Lie algebra
    of the group $SU(\mathcal{H})$).

\item The tangent space $T_{[v]}\mathbb{P}(\mathcal{H})$ can be also
    obtained by pushing forward vectors from $T_{v}\mathcal{H}$ by the
    natural projection $\mathcal{H}\ni v \mapsto [v]\in
    \mathbb{P}(\mathcal{H})$. Let thus
\begin{equation}
\mathbf{x}=\frac{d}{dt}\Big|_{t=0}v(t)\label{TH}
\end{equation}
be the tangent vector to the curve $t\mapsto v(t)\in\mathcal{H}$ at
$v(0)=v$. The corresponding projected curve in $\mathbb{P}(\mathcal{H})$
is given as
\[
t\mapsto\frac{v(t)}{\|v(t)\|}-v\bk{\frac{v}{\|v\|}}{\frac{v(t)}{\|v(t)\|}}.
\]
Differentiating we find the tangent vector $\mathbf{z}$ to
$\mathbb{P}(\mathcal{H})$ at $[v]$, which corresponds to $\mathbf{x}$
\begin{equation}
\label{tangentvP}
\mathbf{z}=\frac{\mathbf{x}}{\|v\|}-
\frac{v}{\|v\|}\bk{\frac{v}{\|v\|}}{\frac{\mathbf{x}}{\|v\|}}.
\end{equation}

\end{enumerate}
Finally, let us remind that the projective space $\mathbb{P}(\mathcal{H})$ is
a K\"ahler manifold (see below). In particular it means it
is a symplectic manifold. The symplectic form at a point
$[v]\in\mathbb{P}(\mathcal{H})$ is given by the formula \cite{GS90}
\begin{gather}
\omega(\hat{\xi}_{1},\hat{\xi}_{2})=-\frac{i\bk v{[\xi_{1}\,,\,\xi_{2}]v}}{2\bk vv},\quad\xi_{1},\,\xi_{2}\in\mathfrak{su}(\mathcal{H}),\label{eq:FS1}
\end{gather}
where $\hat{\xi}_{1}$ and $\hat{\xi}_{2}$ are the fundamental vector fields
associated to $\xi_{1}$ and $\xi_{2}$ \textit{via} (\ref{fundvectf}) (see the
first remark above). At the same time $\mathbb{P}(\mathcal{H})$ is a
Riemannian manifold with the Riemannian metric
\begin{equation}
b(\hat{\xi}_{1},\hat{\xi}_{2})=\frac{\bk{\xi_{1}v}v\bk v{\xi_{2}v}}{\bk vv^{2}}.\label{riemann}
\end{equation}
Both structures are mutually compatible and compatible with the complex
structure, i.e.\
$b(\hat{\xi}_{1},\hat{\xi}_{2})=\omega(\hat{\xi}_{1},i\hat{\xi}_{2})$ which
makes $\mathbb{P}(\mathcal{H})$ a K\"ahler manifold.

Since $\bk{\cdot}{\cdot}$ is $K$-invariant and the symplectic form $\omega$
is given in terms of it, the action of $K$ on $\mathbb{P}(\mathcal{H})$ is
symplectic. This in turn implies that if $K$ is semisimple (which is the case
at hand) then there exists the unique momentum map
$\mu:\mathbb{P}(\mathcal{H})\rightarrow\mathfrak{k}^{\ast}$. Now we are ready
do describe

\paragraph{The construction of the momentum map}

For any $v\in\mathcal{H}$ we define the map
\begin{gather*}
n_{v}:G\rightarrow\mathbb{R},\,\,\, n_{v}(g)=\bk{gv}{gv}=||gv||^{2}
\end{gather*}
The derivative of $n_{v}$ at the group identity, $e\in G$, is by
definition a linear functional $dn_{v}:\mathfrak{g}\rightarrow\mathbb{R}$
and hence an element of $\mathfrak{g}^{\ast}$. Notice however that
since $\bk{\cdot}{\cdot}$ is $K$-invariant, for any $\xi\in\mathfrak{k}$
we have
\begin{gather*}
\langle dn_{v},\,\xi\rangle=
\frac{d}{dt}\Big|_{t=0}\bk{\exp(t\xi)v}{\exp(t\xi)v}=0
\end{gather*}
Next, since $G$ is reductive, its Lie algebra $\mathfrak{g}$ can be
decomposed as $\mathfrak{g}=\mathfrak{k}\oplus i\mathfrak{k}$ and in fact
$dn_{v}\in i\mathfrak{k}^{\ast}$. We define the map
\begin{gather*}
\mu:\mathbb{P}(\mathcal{H})\rightarrow i\mathfrak{k}^{\ast},\,\,\,\mu([v])=\frac{1}{4}\frac{dn_{v}}{\bk vv}.
\end{gather*}
Straightforward calculations give
\begin{gather}
\langle\mu([v]),\, i\xi\rangle=\frac{i}{2}\frac{\bk v{\xi v}}{\bk vv},\,\,\xi\in\mathfrak{k},\label{eq:moment1}
\end{gather}
It is easy to see that if we define $\mu_{i\xi}$ by
$\mu_{i\xi}=\langle\mu([v]),\, i\xi\rangle$ then
$d\mu_{i\xi}=\omega(\hat{\xi},\cdot)$. One can also check that $\mu$ is
equivariant, i.e.
\begin{gather}
\mu([gv])=\mathrm{Ad}_{g}^{\ast}\mu([v]),\,\, g\in K.\label{eq:equiv1}
\end{gather}
In order to avoid confusions we clarify that the Hamilton function associated
to the Hamiltonian vector field $\hat{\xi}$, where $\xi\in\mathfrak{k}$ is
$\mu_{i\xi}$. Equivalently, one can define the true momentum map
$\bar{\mu}_{\xi}=\mu_{i\xi}$ which has all required properties and in
addition satisfies $d\bar{\mu}_{\xi}=\omega(\hat{\xi},\cdot)$. In the
following we will use the former convention.

\subsection{The convexity property of the momentum map}
\label{sub:The-convexity-property}

As the last tools for our applications we need some facts concerning the
convexity property of the momentum map. Let, as previously, $(M,\omega)$ be a
symplectic manifold endowed with the symplectic action of a compact connected
group $K$. Let us fix $T\subset K$, a maximal torus in $K$. As before, the
Lie algebra of $K$ will be denoted by $\mathfrak{k}$ and its Cartan
subalgebra (the Lie algebra of $T$) by $\mathfrak{t}$. Positive elements in
$\mathfrak{t}$ form the so-called positive Weyl chamber $\mathfrak{t_{+}}$
\cite{hall03}. Finally let $\mu:M\rightarrow\mathfrak{k}^{\ast}$ be the
momentum map. Notice that since $\mu$ is equivariant, any $K$ orbit in $M$ is
mapped onto a coadjoint orbit in $\mathfrak{k}^{\ast}$. Moreover, every
coadjoint orbit crosses $\mathfrak{t}_{+}^{\ast}$ in exactly one point. We
can thus define $\Psi:M\rightarrow\mathfrak{t}_{+}^{\ast}$ by
$\Psi(x)=\mu(K.x)\cap\mathfrak{t}_{+}^{\ast}$. By restricting the
identification (\ref{eq:mu star}) to $\mathfrak{t_{+}^{\ast}}$ we may
alternatively treat $\Psi(x)$ as an element of $\mathfrak{t_{+}}$ whenever it
is more convenient.

The following is true \begin{theorem}\label{covexity-theo}The set $\Psi(M)$
is a convex polytope. \end{theorem}

\noindent This theorem has a long history and many people contributed into its
final form. First, Atiyah \cite{Atiyah82} and Guillemin and Sternberg
\cite{GS82} showed that for an abelian $K$ even more is true, namely the whole
$\mu(M)$ is a convex polytope. Guillemin and Sternberg also noticed that for a
non-abelian $K$ the set $\mu(M)$ is typically not convex and proved that
$\Psi(M)$ is a convex polytope when one assumes in addition that $M$ is a
K\"ahler manifold \cite{GS84}. Finally, Kirwan \cite{Kirwan82} showed that
Theorem~\ref{covexity-theo} holds for any connected symplectic manifold. We
will call $\Psi(M)$ the Kirwan polytope.

In this paper we are especially interested in the setting when
$M=\mathbb{P}(\mathcal{H})$ together with the action of a complex reductive
group $G=K^{\mathbb{C}}$. In this case the finer results are known. We give
here one of particular importance for our purposes:

\begin{theorem}\label{(Brion,-Mumford)}Let $G.x$ be an orbit of $G$
through $x\in\mathbb{P}(\mathcal{H})$. Then
\begin{enumerate}
\item The set $\Psi(\overline{G.x})$ is a convex polytope \cite{Ness84},
\cite{brion87}.
\item There is an open dense set of points in $\mathbb{P}(\mathcal{H})$ for
    which $\Psi(\overline{G.x})=\Psi(\mathbb{P}(\mathcal{H}))$
    \cite{GuliSjamaar06}.
\item The collection of different polytopes $\Psi(\overline{G.x})$ where
    $x$ ranges over $M$ is finite \cite{GuliSjamaar06}.
\end{enumerate}
\end{theorem}

\section{The function $||\mu||^{2}$ and its critical sets}
\label{sec:The-function-mu2}

Let us concentrate now on properties of the norm square of
the momentum map. In particular we are interested in the structure of its
critical points and their Morse indices. We start with the physical meaning
of $||\mu||^{2}$. Further, we explicitly compute conditions describing
critical points and finally discuss Morse indices of critical points and
their relevance for the description of entanglement.

For practical reasons it is customary to identify $i\mathfrak{k}^{\ast}$ with
$i\mathfrak{k}$ by means of the $\mathrm{Ad}_{K}$ invariant scalar product
$(\cdot|\cdot)$ on $\mathfrak{k}$. More precisely, for any $\alpha\in
i\mathfrak{k}^{\ast}$ we define $\alpha^{\ast}\in i\mathfrak{k}$ such that
\begin{gather}
\label{eq:mu star} \langle\alpha,\,\xi\rangle=(\alpha^{\ast}|\xi), \,\,\,
\forall\xi\in i\mathfrak{k}.
\end{gather}
To give an explicit formula for $\mu^{\ast}(x)$, where $x=[v]$,
let us choose orthonormal basis $\{\xi_{i}\}$ of $i\mathfrak{k}$.
Then
\begin{gather}
\mu^{\ast}(x)=\sum_{k=1}^{\dim K}(\mu^{\ast}(x)|\xi_{k})\xi_{k}
=\sum_{k=1}^{\dim K}\langle\mu(x),\,\xi_{k}\rangle\xi_{k}
=\sum_{k=1}^{\dim K}\mu_{\xi_{k}}(x)\xi_{k}\label{eq1}.
\end{gather}
After such an identification we define
\begin{gather}
||\mu||^{2}:\mathbb{P}(\mathcal{H})\rightarrow\mathbb{R},\nonumber \\
||\mu||^{2}(x)=(\mu^{\ast}(x)|\mu^{\ast}(x))=
\sum_{k=1}^{\dim K}\mu_{\xi_{k}}^{2}(x)
=\frac{1}{4}\sum_{k=1}^{\dim K}\left(\frac{\bk v{\xi_{k}v}}{\bk vv}\right)^{2}.
\label{eq:norm2}
\end{gather}
Let us note that since $\mu$ is $K$-equivariant the function $||\mu||^{2}$
is $K$-invariant, i.e.
\begin{gather*}
||\mu||^{2}([gv])=||\mu||^{2}([v]),\,\, g\in K.
\end{gather*}

\subsection{Physical meaning of \textmd{$||\mu||^{2}$}}

Remarkably, we can give to $\left\Vert \mu\right\Vert ^{2}$ a clear physical
meaning in the sense that it can be expressed as the mean value of
an observable. We define after Klyachko \cite{klyachko07} the total variance
of a state $[v]\in\mathbb{P}(\mathcal{H})$, with respect to the symmetry
group $K\subset SU(\mathcal{H})$ :
\begin{equation}
\mathrm{Var}([v])=\frac{1}{\bk vv}\left(\sum_{i=1}^{\dim K}\bra v\xi_{i}{}^{2}\ket v
-\frac{1}{\bk vv}\sum_{i=1}^{\dim K}\bra v\xi_{i}\ket v^{2}\right).
\label{eq:variance}
\end{equation}
The above expression is simply the sum of the variances of the observables
$\xi_{i}$ calculated in the state $[v]$. On the other hand,
\begin{gather*}
\mathcal{C}_{2}=\sum_{i=1}^{\dim K}\xi_{i}^{2}
\end{gather*}
is the representation of the second order Casimir operator  which commutes
with every $\xi_{i}$ \cite{BR80}. We can now rewrite
Equation~(\ref{eq:variance2}) in the following manner
\begin{eqnarray}
\mathrm{Var}([v]) & = & \frac{1}{\bk vv}\left(\bra v\mathcal{C}_{2}\ket v
-\frac{1}{\bk vv}\sum_{i=1}^{\dim\, K}\bra v\xi_{i}\ket v^{2}\right).
\label{eq:variance2}
\end{eqnarray}
If the group $K$ acts on $\mathbb{P}(\mathcal{H})$ irreducibly,
which is always the case in our setting, then $\mathcal{C}_{2}$ is
proportional to the identity operator and thus (\ref{eq:variance2})
boils down to
\begin{equation}
\label{eq:var3}
\mathrm{Var}([v])=c-\frac{1}{\bk vv^{2}}\left(\sum_{i=1}^{\dim K}\bra
v\xi_{i})\ket v^{2}\right) =c-4\cdot\left\Vert \mu\right\Vert
^{2}([v]),
\end{equation}
where $c=\frac{\bra v\mathcal{C}_{2}\ket v}{\bk vv}$ is a $[v]$-independent
constant. Notice that since $\left\Vert \mu\right\Vert ^{2}([v])$ is
$K$-invariant, the variance $\mathrm{Var}([v])$ is also $K$-invariant.
Moreover $\mathrm{Var}([v])$ takes the minimal value exactly for separable
states and the maximal value for "`maximally entangled"' states
\cite{klyachko07, oszmaniec12} . It can thus serve as an entanglement measure.
Alternatively $\left\Vert \mu\right\Vert ^{2}([v])$ can be expressed as the
expectation value of $\mathcal{C}_{2}$ represented on the symmetric tensor
product $\mathcal{H}\vee\mathcal{H}$ (see also \cite{oszmaniec12}). Indeed,
one easily checks that for
\begin{equation}
\mathcal{C}_{2}^{\vee}=\sum_{i=1}^{\dim K}(\xi_{i}\otimes I+
I\otimes\xi_{i})^{2},
\label{casimir2}
\end{equation}
which is the second order Casimir operator on $\mathcal{H}\vee\mathcal{H}$
we have
\begin{eqnarray}
\frac{1}{\bk vv^{2}}\bra{v\otimes v}C_{2}^{\vee}\ket{v\otimes v}
=2c+2\frac{1}{\bk vv^{2}}\sum_{i=1}^{\dim K}\bra v\xi_{i}\ket v^{2}
=2c+8\left\Vert \mu\right\Vert ^{2}(\left[v]\right),
\label{eq:step2-1}
\end{eqnarray}
where $c$ is the constant which appeared in (\ref{eq:var3}). Finally, let us
emphasize that by the formula (\ref{eq:var3}) the critical sets of the total
variance of state function are exactly the critical sets of $||\mu||^{2}$. In
the subsequent sections we show how one can use these critical sets to divide
all SLOCC classes of states into disjoint families.

\subsection{Critical points of $||\mu||^{2}$
\label{sub:Critical-points}}

It was noticed around a quarter century ago \cite{Kirwan82,Ness84} that
critical points of $||\mu||^{2}$ play an important role in the classification
of $G=K^{\mathbb{C}}$-orbits in $\mathbb{P}(\mathcal{H})$. In the next two
paragraphs we give the description of the set of critical points of
$||\mu||^{2}$. Let us start with a

\paragraph{Formula for $d||\mu||^{2}(x)$}
>From Equation (\ref{eq:norm2}) we have
\begin{gather}
d||\mu||^{2}(x)=d\left(\sum_{i=1}^{\dim K}\mu_{\xi_{i}}^{2}(x)\right)
=2\sum_{i=1}^{\dim K}\mu_{\xi_{i}}(x)d\mu_{\xi_{i}}
=2\sum_{i=1}^{\dim K}\mu_{\xi_{k}}(x)\omega(\widehat{-i\xi_{k}},\cdot)=
\nonumber \\
2\sum_{k=1}^{\dim K}\omega(\mu_{\xi_{k}}(x)\widehat{-i\xi_{k}},\cdot)
=2\omega(\widehat{-i\sum_{k=1}^{\dim K}\mu_{\xi_{k}}(x)\xi_{k}},\cdot)
=2\omega(\widehat{-i\mu^{\ast}(x)},\cdot)=2d\mu_{\mu^{\ast}(x)}.
\label{eq:dmu_formula}
\end{gather}
Notice also that since the symplectic form $\omega$ is nondegenerate
$d||\mu||^{2}(x)=0$ if and only if $\widehat{-i\mu^{\ast}(x)}=0$.

\paragraph{Two kinds of critical points of $||\mu||^{2}$ }

Using formula (\ref{eq:dmu_formula}) we can divide the set of critical
points of $||\mu||^{2}$ into two disjoint classes
\begin{itemize}
\item The so called minimal critical points for which $\mu(x)=0$. Notice
    that if $\mu(x)=0$, or equivalently $\mu^{\ast}(x)=0$ then of course
    $\widehat{-i\mu^{\ast}(x)}=0$ and hence $d||\mu||^{2}(x)=0$, i.e.\ $x$
    is a critical point. All points in the $\mu^{-1}(0)$ are thus critical.
\item The non-minimal critical points for which $\mu(x)\neq0$ and
    $\widehat{-i\mu^{\ast}(x)}=0$.
\end{itemize}
Let us observe that $[v]$ is a critical point if and only if
\begin{equation}
\mu^{\ast}([v])v=\lambda v\label{eig}
\end{equation}
for some $\lambda\in\mathbb{C}$. Indeed, using (\ref{TH}) for the curve
$t\mapsto\exp(-i\mu^{\ast}([v])t)v$ we obtain $\mathbf{x}= -i\mu^{\ast}([v])v$
and from (\ref{tangentvP})
\begin{equation}
\widehat{-i\mu^{\ast}([v])}=\frac{-i}{\|v\|}\left(\mu^{\ast}([v])v
-\frac{1}{\|v\|^{2}}\bk v{\mu^{\ast}([v])v}\, v\right).
\label{mustar}
\end{equation}
Hence, $\widehat{-i\mu^{\ast}([v])}=0$ implies (\ref{eig}) with
\begin{equation}
\lambda=\frac{1}{\|v\|^{2}}\bk v{\mu^{\ast}([v])v}.\label{lam1}
\end{equation}
Conversely, if (\ref{eig}) is fulfilled then by scalar multiplying
its both sides by $v$ we find that $\lambda$ is given by (\ref{lam1}).
Consequently, upon (\ref{mustar}) we obtain $\widehat{-i\mu^{\ast}([v])}=0$.

\subsection{The Morse index at a critical point }
An important information about the critical point of a function is encoded in
its Morse index, i.e.\ the number of negative eigenvalues of the Hessian at
the point. To calculate it for critical points of $||\mu||^{2}$ let us, as
previously, choose an orthonormal basis $\{\xi_{i}\}$ of $i\mathfrak{k}$. We
already know that
\begin{gather*}
d||\mu||^{2}(x)=d\left(\sum_{i=1}^{\dim K}\mu_{\xi_{i}}^{2}(x)\right)=
2\sum_{i=1}^{\dim K}\mu_{\xi_{i}}(x)d\mu_{\xi_{i}}.
\end{gather*}
Let us assume that $x$ is a critical point of $||\mu||^{2}$. The
Hessian of $||\mu||^{2}(x)$ is the matrix of the second derivatives
(or equivalently the corresponding quadratic form) given by
\begin{gather}
\mathrm{Hess}||\mu||^{2}(x)=2\sum_{k=1}^{\dim K}d\mu_{\xi_{k}}\otimes d\mu_{\xi_{k}}
+2\sum_{k=1}^{\dim K}\mu_{\xi_{k}}\mathrm{Hess}\mu_{\xi_{k}}.
\label{eq:hessian}
\end{gather}
For a minimal critical point $x$, i.e.\ $\mu(x)=0$ we have $\mu_{\xi_{k}}(x)=0$
for all $\xi_{k}$, the formula (\ref{eq:hessian}) simplifies to
\begin{gather*}
\mathrm{Hess}||\mu||^{2}(x)=2\sum_{k=1}^{\dim K}d\mu_{\xi_{k}}\otimes d\mu_{\xi_{k}},
\end{gather*}
and the Morse index, i.e.\ the number of negative eigenvalues of
$\mathrm{Hess}||\mu||^{2}(x)$ is equal to zero, $\mathrm{ind}(x)=0$.

Assume now that $x$ is a non-minimal critical point, i.e.\ $\mu(x)\neq0$ and
$\widehat{-i\mu^{\ast}(x)}=0$. We can always choose an orthonormal basis
$\xi_{i}$ of $i\mathfrak{k}$ in a such a way that
$\xi_{1}=\frac{\mu^{\ast}(x)}{||\mu^{\ast}(x)||}$. This means that all
$\mu_{\xi_{k}}(x)\equiv\langle\mu(x),\,\xi_{k}\rangle=(\xi_{k}|\mu^{\ast}(x))=0$
for $k=2,\ldots,\,\dim K$ and
$\mu_{\xi_{1}}(x)=(\frac{\mu^{\ast}(x)}{||\mu^{\ast}(x)||}|\mu^{\ast}(x))=||\mu||(x)$.
The formula (\ref{eq:hessian}) simplifies to
\begin{gather*}
\mathrm{Hess}||\mu||^{2}(x)=2\sum_{k=1}^{\dim K}d\mu_{\xi_{k}}\otimes
d\mu_{\xi_{k}}+2||\mu||(x)\mathrm{Hess}\mu_{\xi_{1}}.
\end{gather*}
We have following theorem \cite{Ness84}: \begin{theorem}\label{Ness}Assume that
$x$ is a critical point of
$||\mu||^{2}:\mathbb{P}(\mathcal{H})\rightarrow\mathbb{R}$. Then the
following is true
\begin{enumerate}
\item $||\mu||^{2}$ restricted to $G.x$ attains its minimum value at $x$.
\item $||\mu||^{2}$ restricted to $G.x$ attains its minimum value on a
unique $K$-orbit, which is of course the orbit $K.x$ through the
point $x$.
\item If $x\in\mu^{-1}(0)$ then the Morse index $\mathrm{ind}(x)=0$.
\item If $x\notin\mu^{-1}(0)$ then $\mathrm{ind}(x)$ may not be zero.
\end{enumerate}
\end{theorem}

Notice that 3 and 4 have been already proved. The proof of 1 is a direct
consequence of the convexity property of the momentum map (see
Section~\ref{sub:The-convexity-property}). For the proof of 2. see
\cite{Ness84} (Theorem 7.1).

We can decompose the tangent space $T_{x}\mathbb{P}(\mathcal{H})$ as the
direct sum
\begin{gather*}
T_{x}\mathbb{P}(\mathcal{H})=T_{x}G.x\oplus\left(T_{x}G.x\right)^{\bot\omega}.
\end{gather*}
It is possible since $T_{x}G.x$ is a complex and therefore a symplectic
vector space. Hence $T_{x}G.x\cap\left(T_{x}G.x\right)^{\bot\omega}=0$.
Moreover these subspaces are orthogonal with respect to the Riemannian metric
$b(\cdot,\cdot)$ (\ref{riemann}) as they both are stable with respect to
multiplication by $i$. By Theorem~\ref{Ness} we know that the Morse index at
any critical point is determined by the restriction of
$\mathrm{Hess}||\mu||^{2}(x)$ to the subspace
$\left(T_{x}G.x\right)^{\bot\omega}$. Notice next that for any vector
$v\in\left(T_{x}G.x\right)^{\bot\omega}$ we have
\begin{gather*}
d\mu_{\xi_{k}}(v)is=\omega(\widehat{\xi_{k}},\, v)=0,
\end{gather*}
and hence
\begin{gather*}
\left(\mathrm{Hess}||\mu||^{2}(x)\right)|_{\left(T_{x}G.x\right)^{\bot\omega}}
=2||\mu||(x)\mathrm{Hess}\mu_{\xi_{1}}.
\end{gather*}
Summing up we proved
\begin{theorem}
The Morse index at the critical point is given by the number of the negative
eigenvalues of $\mathrm{Hess}\mu_{\xi_{1}}$ where
$\xi_{1}=\frac{\mu^{\ast}(x)}{||\mu^{\ast}(x)||}$.
\end{theorem}
Notice that from Theorem \ref{Ness} we can read off a nice
interpretation of the Morse index computed at some critical point. Namely the
Morse index equals the number of non-SLOCC directions at a given critical
point, in which entanglement (measured by $-||\mu||^{2}$) increases.

\section{Stratification of the complex projective space by critical sets of
$||\mu||^{2}$}
\label{sec:stratification}

As it was already discussed in the introduction the space of orbits of SLOCC
action, $\mathbb{P}\mathcal{H}/G$, is in general not a Hausdorff space as
orbits of $G$ are not necessary closed. It is thus natural to overcome this
difficulty by exhibiting some `good' quotient space. In
this section we present two known constructions of such
quotients. First is the so called \emph{categorical quotient}.
The second one, the \emph{Ness stratification}, is in a
sense a generalization of the former. They are both directly related to the
structure of critical sets of $||\mu||^2$  on $\mathbb{P}\mathcal{H}$ . The
categorical quotient uses minimal critical points whereas the Ness
stratification uses both minimal and non minimal critical points. In the
following sections we will discuss the physical interpretation of the Ness
quotient. It steams directly from relevance of $||\mu||^2$ as an entanglement
measure.

\subsection{The minimal critical points and the construction of the
categorical quotient} \label{GIT-construction}

The minimal critical points of $||\mu||^{2}$ play an important role
in the construction of the $G$-orbit space in $\mathbb{P}(\mathcal{H})$.
For simplicity we begin with the construction of the $G$-orbits space
on the level of $\mathcal{H}$.
%
%
First, for any reductive group the following fundamental theorem holds

\begin{theorem}(Hilbert, Nagata \cite{M03}),\label{(Hilbert,-Nagata-)}
If $G$ is reductive, then the ring of $G$-invariant polynomials
$\mathbb{C}[\mathcal{H}]$, i.e.\
polynomials which are constant on the $G$-orbits is finitely generated.
\end{theorem}

Having Theorem \ref{(Hilbert,-Nagata-)} one can ask if it is possible to
distinguish between all $G$-orbits in $\mathcal{H}$ using invariant
polynomials. Recall that for a compact group action all orbits are closed and
therefore the answer is positive. However, for non-compact group $G$ it is
not the case. It is because $G$-invariant polynomials are continuous and
therefore take the same value on orbits for which the intersection of
closures in not empty,
$\overline{G.v_{1}}\cap\overline{G.v_{2}}\neq\emptyset$. We call two orbits
$G.v$ and $G.u$ \emph{closure equivalent} if there exists a sequence of
orbits $G.v_{1}=G.v,\, G.v_{2},\ldots,\, G.v_{n}=G.u$ such that
$\overline{G.v_{k}}\cap\overline{G.v_{k+1}}\neq\emptyset$. This equivalence
relation divides $G$-orbits into closure equivalence classes. The proper
question is hence the following one. Do $G$-invariant polynomials separate
the closure inequivalent orbits? We have the following \cite{M03}

\begin{theorem}Let $G.v_{1}$ and $G.v_{2}$ be two orbits such that
$\overline{G.v_{1}}\cap\overline{G.v_{2}}=\emptyset$. Then there
is a $G$-invariant polynomial which separates $G.v_{1}$ from $G.v_{2}$.

\end{theorem}

The immediate corollary from this theorem is

\begin{corollary} \label{closed-orbit}The following are true:
\begin{enumerate}
\item There is enough $G$-invariant polynomials to separate closed $G$-orbits.
\item Closure equivalence class of orbits can contain at most one closed
orbit, as if it contained more than one then invariant polynomials
would not be constant on it.
\item Every closure equivalence class of orbits contains exactly one closed
orbit (the orbit of lowest dimension). Moreover, it is contained in
the closure of every orbit in the equivalence class.
\end{enumerate}
\end{corollary}

The last point implies that we can simplify the definition of the equivalence
of two orbits $G.v$ and $G.u$ to
$\overline{G.v}\cap\overline{G.u}\ne\emptyset$.

The correct quotient construction is thus with respect to closed orbits. In
the case of the projective space $\mathbb{P}(\mathcal{H})$ one first removes
the so-called \emph{null cone} $N$ which is the closure equivalence class of
orbits which contains $0\in\mathcal{H}$ as the unique closed $G$-orbit. What
is left is denoted by $\mathbb{P}(\mathcal{H})_{ss}$ and called the set of
\emph{semistable} points. Then we say that two points
$x_{1}=[v_{1}]\in\mathbb{P}(\mathcal{H})_{ss}$ and
$x_{2}=[v_{2}]\in\mathbb{P}(\mathcal{H})_{ss}$ are equivalent,
$[v_{1}]\sim[v_{2}]$, if and only if
$\overline{G.v_{1}}\cap\overline{G.v_{2}}\neq\emptyset$. The resulting
quotient space, denoted by $\mathbb{P}(\mathcal{H})_{ss}\sslash G$, is known
in the literature as the \emph{categorical quotient}. Notice that
$\mathbb{P}(\mathcal{H})_{ss}\sslash G$ is parametrized by closed $G$-orbits.
Remarkably, the minimal critical points of $||\mu||^{2}$ parameterize all
closed $G$-orbits in $\mathbb{P}(\mathcal{H})_{ss}$. The
following was proved by George Kempf and Linda Ness \cite{KN82}

\begin{theorem}\label{minimalKep-Ness-1}Assume that $[v]\in\mu^{-1}(0)$,
i.e.\ $[v]$ is a minimal critical point of
$||\mu||^{2}:\mathbb{P}(\mathcal{H})\rightarrow\mathbb{R}$.
Then the orbit $G.v$ is closed and $0\notin\overline{G.v}$. Moreover
all closed $G$-orbits in $\mathbb{P}(\mathcal{H})_{ss}$ are of this
type.
\end{theorem}

It is now easy to see that we have the following isomorphism of quotients
\begin{gather*}
\mu^{-1}(0)/K\cong\mathbb{P}(\mathcal{H})_{ss}\sslash G\,,
\end{gather*}
Moreover, we have \cite{Ness84}

\begin{theorem}\label{proj-mu}The sets $\mathbb{P}(\mathcal{H})_{ss}\sslash G$
and hence $\mu^{-1}(0)/K$ are projective varieties.

\end{theorem}

By Theorem~\ref{proj-mu} one can view the categorical quotient construction
in terms of fibers of certain surjective map between complex spaces, i.e.\ we
define
\begin{gather*}
\pi:\,\mathbb{P}(\mathcal{H})_{ss}\rightarrow\mathbb{P}(\mathcal{H})_{ss}\sslash
G\,,
\end{gather*}
where each fiber of $\pi$ is a closure equivalence class of orbits containing
exactly one closed $G$-orbit which in turn contains the $K$-orbit from
$\mu^{-1}(0)$. It is natural to ask when a fiber of $\pi$ is given by a
single $G$-orbit, i.e.\ when a closure equivalence class of orbits is simply
one closed $G$-orbit. In order to address this question we need the following
general theorem, describing surprising behavior of fiber dimensions of a map
between complex spaces \cite{Milne05}.

\begin{theorem}\label{Alan}Let $\pi:\, X\rightarrow Y$ be a surjective regular
map between complex spaces and assume that each fiber of $\pi$ is connected.
Let $d=\dim X-\dim Y$. The following are true
\begin{enumerate}
\item There exists Zariski open dense set $Y_{reg}\subset Y$ such that
    for $y\in Y_{reg}$ the fiber $\pi^{-1}(y)$ has dimension
    $\dim\pi^{-1}(y)=d$.
\item For each $y\in Y\setminus Y_{reg}$ the dimension of fiber
    $\pi^{-1}(y)$ cannot drop down, i.e. $\dim\pi^{-1}(y)> d$.
\end{enumerate}
\end{theorem}

On the other hand, it is known that \cite{Mumford77}

\begin{theorem}The set $\mathbb{P}(\mathcal{H})_{s}=\{G.x:\,\dim G.x=
\dim G\,\,\mbox{and}\,\,\mbox{\ensuremath{x\in\mu^{-1}(0)}}\}$
if exists is Zariski open dense subset of $\mathbb{P}(\mathcal{H})_{ss}$.

\end{theorem}

The points from $\mathbb{P}(\mathcal{H})_{s}$ are called
\emph{stable points}. Notice now that if $x\in\mathbb{P}(\mathcal{H})_{s}$
then the fiber of $\pi:\,\mathbb{P}(\mathcal{H})_{ss}
\rightarrow\mathbb{P}(\mathcal{H})_{ss}\sslash G$ is exactly $G.x$. To see it
let us assume on the contrary that there is a point
$y\in\mathbb{P}(\mathcal{H})_{ss}$ such that $G.y$ is closure equivalent to
$G.x$. By Corollary~\ref{closed-orbit} this means that either $G.y=G.x$ or
$\dim G.y>\dim G.x$. But the later is impossible since $\dim G.x=\dim G$.
Therefore the generic dimension of a categorical quotient fiber in the
presence of stable points is $d=\dim G$. Combining this observation with
Theorem~\ref{Alan} we arrive at

\begin{theorem}For a semistable but not stable point $x\in\mu^{-1}(0)$
the fiber of categorical quotient $\pi^{-1}(\pi(x))$ contains more than one
$G$-orbit.

\end{theorem}


Finally, let us point out that it is known (see \cite{HH96}) that the set
$G.\mu^{-1}(0)$ is either empty or open and dense in
$\mathbb{P}(\mathcal{H})$. This means that $\mu^{-1}(0)/K$ parameterizes
almost all $G$-orbits in $\mathbb{P}(\mathcal{H})$. Moreover, all these
points are in one-to-one correspondence with the minimal critical sets of
$||\mu||^{2}$.

A natural question now is whether it is possible to stratify the null cone $N$ \emph{via} critical sets of $\left\Vert
\mu\right\Vert ^{2}$. In other words we want to introduce an additional
refinement in the null cone $N$ stemming from the critical sets of the total
variance restricted to it. The resulting quotient is called Ness stratification.

\begin{figure}[H]
~~~~~~~~~~~~~~~~~~~~~~~~~~\includegraphics[scale=0.17]{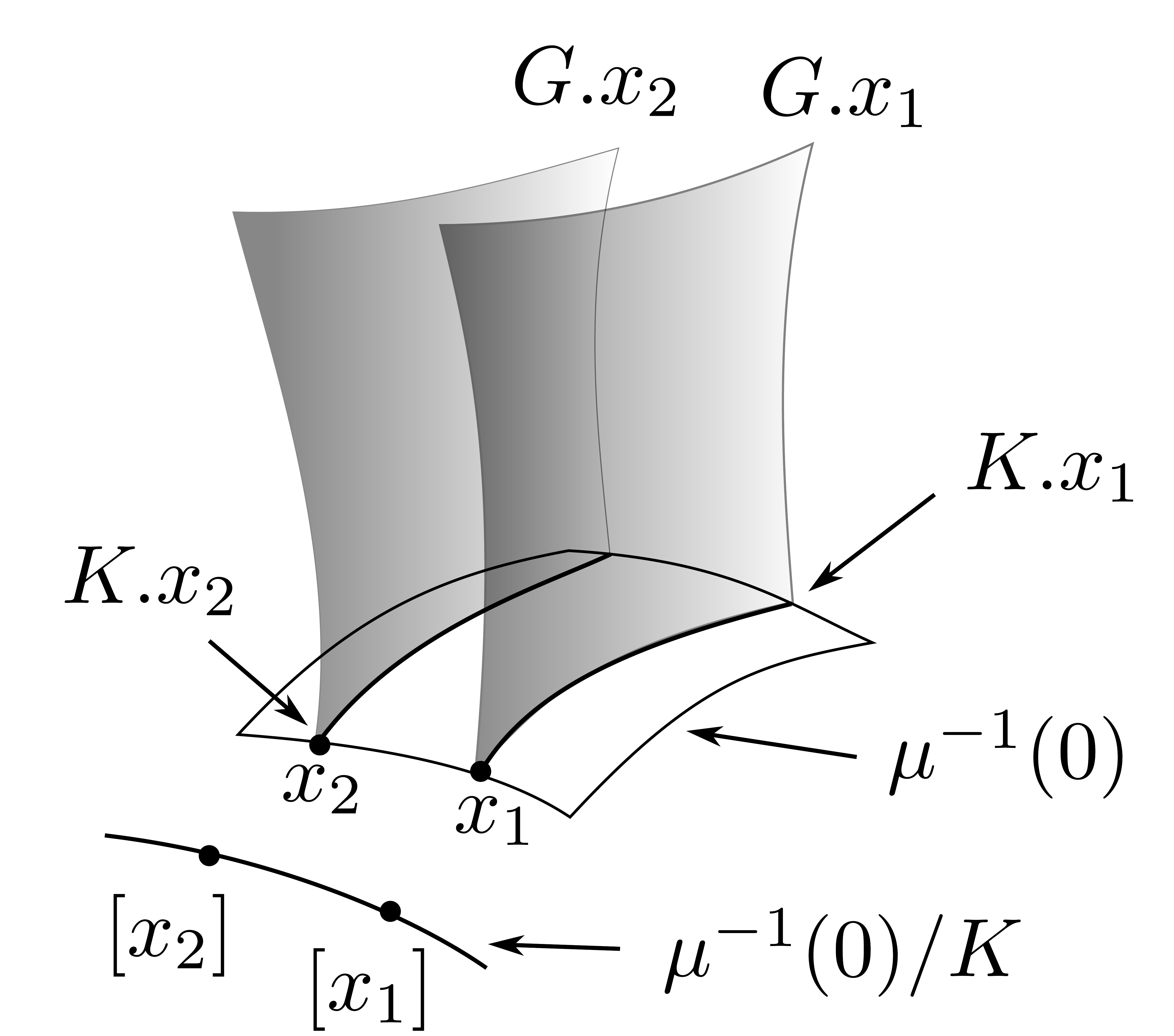}

\caption{The idea of the categorical quotient  construction, i.e.
$\mu^{-1}(0)/K\cong\mathbb{P}(\mathcal{H})_{ss}\sslash G$.}
\end{figure}

\subsection{Stratification of the null cone by critical sets of $||\mu||^{2}$
\label{sub:Startification}}

We are now ready to describe the stratification of the null cone $N$ by
critical sets of $||\mu||^{2}$ and provide the method for finding groups of
families of $G$-orbits. Let us first notice that since Hamiltonian vector
field corresponding to $d\left(-||\mu||^{2}\right)$ is given by
$2\widehat{i\mu^{\ast}(x)}$ and $\mathbb{P}(\mathcal{H})$ is a K\"ahler
manifold, the gradient of $||\mu||^{2}$ with respect to the Riemannian metric
$b(\cdot,\cdot)$ is given by $2\widehat{\mu^{\ast}(x)}$. Next, the vector
field $\widehat{\mu^{\ast}(x)}$ is tangent to the $G$-orbit through $x$ as
$\mu^{\ast}(x)\in i\mathfrak{k}\subset\mathfrak{g}$. Thus any point $x\in N$
is carried by the gradient flow into some critical $K$-orbit of $||\mu||^{2}$
and in this way we obtain stratification of $N$. Recall that similarly every
semistable point in $\mathbb{P}(\mathcal{H})_{ss}$ is taken by the gradient
flow to minimal critical points. Moreover, all minimal critical points
constitute the fiber $\Psi^{-1}(0)$ of the momentum map. It is therefore
natural to divide remaining critical points in $N$ by fibers of $\Psi$. More
precisely for each $\alpha$ such that $\alpha=\Psi(x)$ with $x\in N$, we
check if $\Psi^{-1}(\alpha)$ contains $||\mu||^{2}$-critical $K$-orbits. We
denote by $C_{\alpha}$ the set of $||\mu||^{2}$-critical points such that
$\Psi(K.x)=\alpha$ and by $N_{\alpha}$ all points in $N$ for which
$C_{\alpha}$ is the limit set of the gradient flow. It was shown by Kirwan
\cite{Kirwan82} that $N_\alpha$ is $G$-invariant. Next we say that two points
in $N_{\alpha}$ are $G$-equivalent if and only if they are carried by
gradient flow to the same critical $K$-orbit in $C_{\alpha}$. Under this
equivalence we get the desired stratification of $N$. It turns out that the
set $C_{\alpha}/K$ has a nice structure, namely \cite{Ness84}
\begin{theorem}The set $C_{\alpha}/K$ is a projective variety.
\end{theorem}

\begin{figure}[h]
~~~~~~~~~~~~~~~~~~~~~~~~~~~~~\includegraphics[scale=0.11]{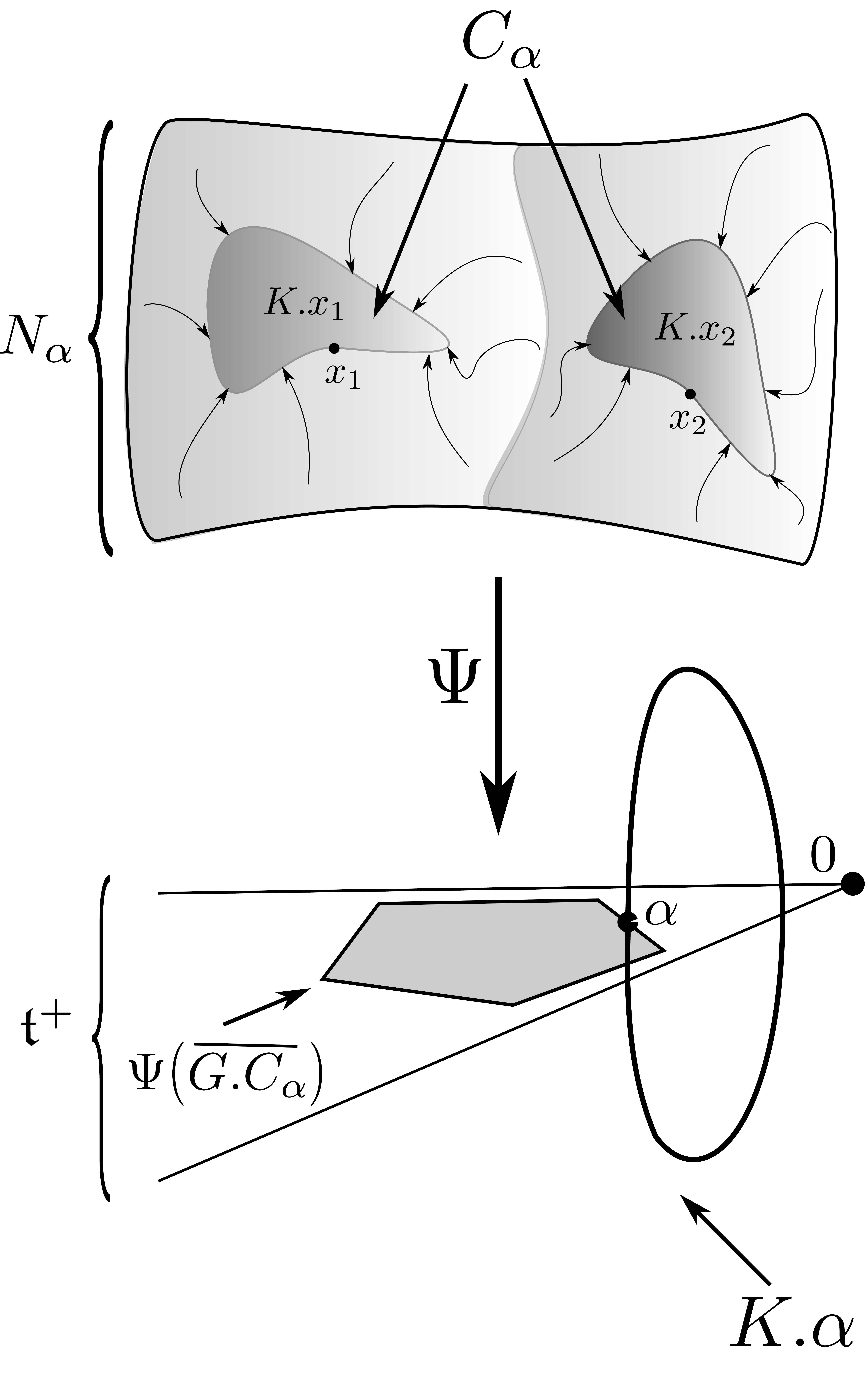}

\caption{The sets $N_\alpha$ and $C_\alpha$, with two exemplary critical $K$-orbits, $K.x_1$ and $K.x_2$. The arrows represent the gradient flow of $-||\mu||^{2}$.}
\end{figure}
In this way the quotient of $\mathbb{P}(\mathcal{H})$ by $G$ is
decomposed into finite number of disjoint projective varieties
\begin{gather*}
\mathbb{P}(\mathcal{H})/G=\bigcup_{\alpha}C_{\alpha}/K
\end{gather*}
The main one is given by $C_{0}/K\simeq\mu^{-1}(0)/K$ or equivalently
$\Psi^{-1}(0)/K$. The other are $C_{\alpha}/K$ where $C_{\alpha}$ is the set
of all critical $K$-orbits in $\Psi^{-1}(\alpha)$. The only unclear fact
concerns finite rather than infinite number of these varieties. It can be
easily clarified using point 3 of Theorem \ref{(Brion,-Mumford)}. To this end
note that one can equivalently define $N_\alpha$ as all points $x\in
\mathbb{P}(\mathcal{H})$ for which polytopes $\Psi(\overline{G.x})$ have the
same closest point to the origin. As the number of polytopes is finite the
number of varieties $C_\alpha$ is as well finite.

\paragraph{An algorithm for finding classes of $G$-orbits}

Summing up, in order to find the above described decomposition of
$\mathbb{P}(\mathcal{H})$ one has to
\begin{enumerate}
\item Find the set $\mu^{-1}(0)/K$. Usually it is a hard problem and
    requires knowledge of a canonical form of $x\in\mathbb{P}(\mathcal{H})$
    under $K$-action.
\item Find sets $C_{\alpha}/K$ where $C_{\alpha}$ is the set of all critical
$K$-orbits in $\Psi^{-1}(\alpha)$. This happens to be relatively
easy task as we show in the next section.
\end{enumerate}

In the next sections we discuss how this algorithm and other above described
ideas can be used for better understatement of multipartite entanglement. The
goal is obtained by calculating various examples. For clarity we now give a
table which should be treated as a dictionary between the abstract concepts
and their QI counterparts. The contents of the table is also discussed in the
subsequent sections.
\begin{table}[h]
  \centering
  \caption{A dictionary }\label{table}
    \begin{tabular}{|p {0.25\linewidth}|p {0.6\linewidth}|}
    \hline
    $G$-orbit & SLOCC class of states \tabularnewline
    \hline
    the momentum map $\mu$ & the map which assigns to the state $[v]$ the collection of its reduced one-particle density matrices\tabularnewline
    \hline
    $||\mu||^2([v])$ & the total variance of state $\mathrm{Var}([v])$ \tabularnewline
    \hline
    closure equivalence class of orbits & family of asymptotically equivalent SLOCC classes  \tabularnewline
    \hline
    stable point & SLOCC family consists of exactly one SLOCC class \tabularnewline
    \hline
    semistable but not stable point & SLOCC family consists of many SLOCC class \tabularnewline
    \hline
    $\Psi(\overline{G.[v]})$ & SLOCC momentum polytope, collection of all possible spectra of reduced one-particle density matrices for $[u]\in\overline{G.[v]} $ \tabularnewline
    \hline
    strata $N_\alpha$ & group of families of SLOCC classes - all states for which SLOCC momentum polytopes have the same closest point to the origin\tabularnewline
    \hline
    $C_\alpha$ &  set of critical points of $\mathrm{Var}([v])$ with the same spectra of reduced one-particle density matrices \tabularnewline
    \hline
  \end{tabular}
\end{table}

\section{The SLOCC classes for distinguishable particles
\label{sec:The-SLOCC-invariant}}

We are now ready to apply the mathematical tools described in sections
2 and 3 to the characterization of SLOCC classes of states.

\paragraph{Entanglement in systems of distinguishable particles}

In the following we put
$\mathcal{H}=\mathbb{C}^{N}\otimes\ldots\otimes\mathbb{C}^{N}$ which is the
Hilbert space of $L$ identical but distinguishable particles. The group $G$
is a direct product of $L$ copies of $SL(N,\mathbb{C})$, i.e.\
$G=SL(N,\mathbb{C})^{\times L}$ and $K=SU(N)^{\times L}$. We have
$G=K^{\mathbb{C}}$ and hence the group $G$ is reductive. Moreover, $K$ is
semisimple compact and connected. As mentioned in the Introduction the group
$G$ represents the so-called SLOCC operations whereas $K$ stands for local
unitary operations. The actions of both $G$ and $K$ on $\mathcal{H}$ are
given by
\begin{gather*}
(U_{1},\, U_{2},\ldots,\, U_{L}).v=U_{1}\otimes U_{2}\otimes
\ldots\otimes U_{L}v.
\end{gather*}
The Lie algebras of $G$ and $K$ are direct sums of $L$ copies of
$\mathfrak{sl}(N)$ and $\mathfrak{su}(N)$ respectively, i.e.\
$\mathfrak{g}=\mathfrak{sl}(N)^{\oplus L}$ and
$\mathfrak{k}=\mathfrak{su}(N)^{\oplus L}$. The corresponding actions of
$\mathfrak{g}$ and $\mathfrak{k}$ on $\mathcal{H}$ read
\begin{gather*}
(\xi_{1},\,\xi_{2},\ldots,\,\xi_{L})v=(\xi_{1}\otimes I_{N}
\otimes\ldots\otimes I_{N}
+I_{N}\otimes\xi_{2}\otimes\ldots\otimes I_{N}+\ldots+\\
+\ldots+I_{N}\otimes\ldots\otimes\xi_{L}v),
\end{gather*}
where $I_{N}$ is the $N\times N$ identity matrix. The momentum map
is $\mu:\mathbb{P}(\mathcal{H})\rightarrow i\mathfrak{k}$ is given
by \cite{sawicki11}
\begin{gather*}
\mu([v])=(\rho_{1}([v])-\frac{1}{N}I_{N},\,\rho_{2}([v])
-\frac{1}{N}I_{N},\ldots,\rho_{L}([v])-\frac{1}{N}I_{N}),
\end{gather*}
where $\rho_{i}([v])$ is the $i$-th reduced single-particle density matrix of
the state $\frac{v}{||v||}$. The points $[v]\in\mathbb{P}(\mathcal{H})$ for
which $\mu([v])=0$ are thus precisely the states for which all reduced
single-particle density matrices $\rho_{i}$ are maximally mixed. Moreover the
map $\Psi:\mathbb{P}(\mathcal{H})\rightarrow\mathfrak{it}_{+}$ is given by
\begin{gather*}
\Psi([v])=(\tilde{\rho}_{1}([v])-\frac{1}{N}I_{N},\,\tilde{\rho}_{2}([v])
-\frac{1}{N}I_{N},\ldots,\tilde{\rho}_{L}([v])-\frac{1}{N}I_{N}),
\end{gather*}
where $\tilde{\rho}_{i}([v])=\mbox{diag}\left(p_{1}^{i},
\ldots,p_{N}^{i}\right)$,
$p_{k}^{i}\ge p_{k+1}^{i}$, and $\mbox{Spect}(\rho_{i}([v]))
=\left\{ p_{1}^{i},\ldots,p_{N}^{i}\right\} $.
In what follows we will need the explicit formula for $\mu^{\ast}$:
\begin{gather}
\mu^{\ast}([v])=\left(\rho_{1}([v])-\frac{1}{N}I_{N}\right) \otimes
I_{N}\otimes\ldots\otimes I_{N}+\nonumber \\
+ I_{N}\otimes\left(\rho_{2}([v])
-\frac{1}{N}I_{N}\right)\otimes I_{N}\otimes\ldots\otimes I_{N}+\nonumber \\
+\ldots+I_{N}\otimes\ldots\otimes I_{N}\otimes\left(\rho_{2}([v])
-\frac{1}{N}I_{N}\right)\,.
\end{gather}
The image of $\Psi$ is by Theorem~\ref{covexity-theo} a convex polytope. For
the system of $L$ qubits, i.e.\ when $N=2$ this polytope is given by
particularly simple inequalities, namely we have the following theorem
\cite{Higuchi03}

\begin{theorem}\label{polytope}For a $L$-qubit system all constraints
on the $1$-qubit reduced density matrices of a pure state are given
by polygonal inequalities
\begin{gather*}
p_{i}\leq\sum_{j\neq i}p_{j}
\end{gather*}
for the minimal eigenvalues $p_{i}$ of $\rho_{i}$. \end{theorem}

Let us now give some details about finding SLOCC classes for identical
distinguishable particles. First we have the following theorem which gives
canonical form of the state up to local unitaries \cite{VDM03}:

\begin{theorem}\label{thm1}For a general state
$\Psi=\sum C_{i_{1},\ldots,i_{L}}\ket{i_{1}\ldots i_{L}}$
in $\mathcal{H}=\mathbb{C}^{N}\otimes\ldots\otimes\mathbb{C}^{N}$
there exist local unitaries $U_{i}$ such that all the following entries
in the state $\Psi^{\prime}=U_{1}\otimes\ldots\otimes U_{L}\Psi$
are set to zero:
\begin{gather*}
\forall1\leq j\leq N\,\,\forall k>j\,\,\, C_{jj\ldots jk}^{\prime}
=C_{jj\ldots jkj}^{\prime}=\ldots=C_{kj\ldots jj}^{\prime}=0.
\end{gather*}
Moreover all entries $C_{NNiN\ldots N}^{\prime}$, where $i\leq N$ can be made
real and positive. If the number of parties exceeds $2$, then the normal form
is typically not unique up to permutations, but there exist a discrete number
of different normal forms with the aforementioned property. \end{theorem}

The algorithm from Section~\ref{sub:Startification} thus reads
\begin{enumerate}
\item Use Theorem \ref{thm1} to find $\mu^{-1}(0)/K$. This gives the main
    group of families of SLOCC classes in $\mathbb{P}(\mathcal{H})_{ss}$.
    We show in next section how to do it explicitly for two, three and
    four qubits.
\item For the groups of families of SLOCC classes in the null cone $N$ we
    have the following strategy. Since critical sets of $||\mu||^{2}$ are
    $K$-invariant it is enough to check when $\mu^{\ast}([v]).v=\lambda
    v$ for
    $\mu^{\ast}([v])\in\Psi(\mathbb{P}(\mathcal{H}))\setminus\{0\}$. In
    other words we have to go over the Kirwan polytope and verify for
    which states $v$ we have $\alpha^{\ast}v=\lambda v$,
    $\alpha^{\ast}\in\Psi(\mathbb{P}(\mathcal{H}))\setminus\{0\}$. For
    many-qubit systems this polytope is described by polygonal
    inequalities (see Theorem \ref{polytope}) which are explicitly known.
    Moreover, we can treat $\alpha^{\ast}$ as an $N^{L}\times N^{L}$
    diagonal matrix acting on $\mathcal{H}$. If the diagonal elements of
    $\alpha^{\ast}$ are nondegenerate the corresponding eigenvectors are
    separable states. In order to find other nontrivial SLOCC classes we
    are interested in the situation when the spectrum of $\alpha^{\ast}$
    is degenerate. Remarkably, the dimensions of the degenerate
    eigenspaces are typically relatively small. For consistency, for each
    of such eigenspaces we choose these vectors $v$ for which
    $\alpha=\mu\left(\left[v\right]\right)$. In this way we arrive at
    sets $X_{\alpha}=\mu^{-1}(\alpha)\cap C_{\alpha}$, where
    $\alpha=\Psi([v])$. In order to obtain set $C_{\alpha}/K$ we still
    need to get rid of the remaining $K$ freedom. We do it by
    constructing the space $C_{\alpha}/K=X_{\alpha}/K_{\alpha}$ where
    $K_{\alpha}$ is the isotropy subgroup of $\alpha$ with respect to the
    adjoint action of $K$. In the next section we show how to do it for
    two distinguishable particles as well as for three and four qubits.
\end{enumerate}
Before we give some details about application of the algorithm described
in Section~\ref{sub:Startification} let us describe additional ideas
connecting critical points of $||\mu||^{2}$ and the convexity property
of the momentum map with entanglement.
\begin{enumerate}
\item To any point $[v]\in\mathbb{P}(\mathcal{H})$ we can associate a
    positive number in the following way. For
    $[v]\in\mathbb{P}(\mathcal{H})$ consider the convex polytope
    $\Psi(\overline{G.[v]})$. We put
    $d([v])\equiv\mbox{dist}(0,\,\Psi(\overline{G.[v]}))$, i.e.\ $d([v])$
    is the distance from $0$ to the polytope $\Psi(\overline{G.[v]})$.
    Since $d([v])=\mbox{inf}_{y\in G.[v]}||\mu(y)||$ we have $d([v])=0$ for
    $[v]\in\mu^{-1}(0)$. Moreover, since $0\in\Psi(\overline{G.x})$, this
    distance is also zero for any $x\in G.\mu^{-1}(0)$. In other words the
    states which can be transformed by SLOCC operations to a state for
    which all one particle density matrices are maximally mixed are
    characterized by $d([v])=0$. Hence $d([v])\neq0$ if and only if $[v]$
    is in the null cone $N$.
\item Using the results described in Section~\ref{sec:The-function-mu2} we
    get a nice characterization of points $x\in G.[v]$ for which
    $||\mu^{\ast}(x)||=d([v])$. By Theorem~\ref{Ness} we know that these
    are points from the critical $K$-orbits of $||\mu||^{2}$. Moreover by
    definition of the total variance of a state $\mathrm{Var}[v]$, the
    states belonging to this critical $K$-orbit are the most entangled
    representatives of the appropriate SLOCC family.
\item The Morse index at any point $[v]$ belonging to the critical
    $K$-orbit equals to the number of independent directions representing
    the (non-SLOCC) directions in which the variance $\mathrm{Var}[v]$, or
    equivalently entanglement, increases. Notice that by Theorem~\ref{Ness}
    for states in $\mu^{-1}(0)$ this index is always $0$, i.e.\ even
    non-SLOCC operations can not increase the total variance of states
    belonging to $\mu^{-1}(0)$. On the other hand for a separable state
    $[v]$ the Morse index is maximal and equal to
    $\dim\mathbb{P}(\mathcal{H})-\dim G.[v]$. It is simply because any
    non-SLOCC operation increases entanglement of separable states.
\end{enumerate}
Summing up, both $d([v])$ and the Morse index have an interesting
geometric and physical interpretations and can be considered as entanglement
measures. Notice finally that by equation (\ref{eq:var3}) which relates
$\mathrm{Var}[v]$ and $||\mu||^{2}([v])$ we can rephrase $d([v])$
in terms of square root of the total variance
\begin{gather*}
d([v])=\frac{1}{2}\sqrt{c-\mbox{sup}_{u}\mathrm{Var}([u])},
\end{gather*}
where the maximum is taken over the states $[u]$ belonging to the SLOCC class
of state $[v]$. Finally, let us clarify that by the above described procedure
we divide the set of pure states $\mathbb{P}(\mathcal{H})$ into three
disjoint classes. The first one contains states which can be converted by $G$
to states characterized by maximally mixed reduced density matrices. The
second one posses the same property only asymptotically and the third one
does not have it. Moreover, the first class, if exists, contains almost all
states from $\mathbb{P}(\mathcal{H})$.

\section{The main examples for distinguishable particles}
\label{sec:The-main-examples}

We illustrate the ideas described in Section~\ref{sec:The-SLOCC-invariant} by
calculations for bipartite states and three and four qubits. We show that for
bipartite and three qubits all SLOCC classes contain critical sets of the
total variance of state. Moreover, the set of semistable
points consists only of stable points. The families of SLOCC classes are
given buy single $G$-orbits and they agree with these known in the
literature. For four qubits it is not the case. We show that only for generic
states from the so-called $G_{abcd}$  class \cite{VDDV02} the families of
SLOCC classes are given by single $G$-orbits. We do have states which are
semistable but not stable. Almost all classes found in \cite{VDDV02} are of
this type, i.e. they are closure equivalent to non-generic states from
$G_{abcd}$. In all cases we calculate the Morse indices at critical sets of
the total variance of state and the value of our geometric SLOCC invariant
measure of entanglement.

\subsection{Bipartite case}\label{bipartite}

Let us consider two identical but distinguishable particles, i.e.\ the
Hilbert space of the system is
$\mathcal{H}=\mathbb{C}^{N}\otimes\mathbb{C}^{N}$. The SLOCC operations are
elements of the group $G=SL(N,\mathbb{C})\times SL(N,\mathbb{C})$ and local
unitary operations of the group $K=SU(N)\times SU(N)$. It is well known that
up to a $K$-action any state $v\in\mathcal{H}$ can be written in the
so-called Schmidt form, i.e.\
\begin{gather}
v=\sum_{i=1}^{N}a_{i}\ket i\otimes\ket i\equiv\sum_{i=1}^{N}a_{i}\ket{i,i},
\label{eq:schmidt}
\end{gather}
where $\{\ket i\}_{i=1}^{N}$ is an orthonormal basis in $\mathcal{H}$,
$a_{i}\geq0$ and $\sum_{i=1}^{N}a_{i}^{2}=1$. Our first task is to determine
the critical points of $||\mu||^{2}$. From Section~\ref{sub:Critical-points} we
know that $[v]\in\mathbb{P}(\mathcal{H})$ is a critical point of $||\mu||^{2}$
if and only if
\begin{gather*}
\widehat{\mu^{\ast}([v])}=0\,\Leftrightarrow\,\mu^{\ast}([v])v=\lambda v,
\end{gather*}
for some $\lambda\in\mathbb{C}$, i.e.\ when $v$ is an eigenvector
of
\begin{gather*}
\left(\left(\rho_{1}([v])-\frac{1}{N}I_{N}\right)\otimes
I_{N}+I_{N}\otimes\left(\rho_{2}([v])-\frac{1}{N}I_{N}\right)\right)v
\end{gather*}
where $\rho_{i}([v])$ are the reduced one-particle density matrices.
Notice that since $||\mu||^{2}$ is $K$-invariant it is enough to
consider states $v$ in the Schmidt form, which we assume in what
follows. For the state (\ref{eq:schmidt}) we have
\begin{gather*}
\rho_{1}([v])=\rho_{2}([v])=\mbox{diag}\left(a_{1}^{2},\, a_{2}^{2},\ldots,\,
a_{N}^{2}\right),
\end{gather*}
and
\begin{gather*}
\mu^{\ast}([v])v=\sum_{i=1}^{N}\left(a_{i}^{2}-\frac{1}{N}\right)a_{i}\ket{i,i}.
\end{gather*}
To verify when
\begin{gather}
\sum_{i=1}^{N}\left(a_{i}^{2}-\frac{1}{N}\right)a_{i}\ket{ii}=
\lambda\sum_{i=1}^{N}a_{i}\ket{i,i},\label{eq:condition1}
\end{gather}
we consider two cases
\begin{enumerate}
\item All Schmidt coefficients $a_{1},\ldots,a_{N}\neq0$. In this case the
    condition (\ref{eq:condition1}) implies that
\begin{gather*}
a_{i}^{2}-\frac{1}{N}=\lambda,\,\,\mbox{for all}\, i\in\{1,\ldots,\, N\}.
\end{gather*}
Since $a_{i}\geq0$ and $\sum_{i=1}^{N}a_{i}^{2}=1$ it is possible
if and only if $\lambda=0$, i.e.\ $a_{i}=\frac{1}{\sqrt{N}}$ for
all $i\in\{1,\ldots,\, N\}$.
\item The coefficients $p_{k+1},\ldots,p_{N}=0$, where $k\geq1$. In this
    case the condition (\ref{eq:condition1}) imposes
\begin{gather*}
a_{i}^{2}-\frac{1}{N}=\lambda,\,\,\mbox{for all}\, i\in\{1,\ldots,\, k\},
\end{gather*}
which is possible if and only if $\lambda=\frac{N-k}{Nk}$, i.e.\
$p_{i}=\frac{1}{\sqrt{k}}$ for all $i\in\{1,\ldots,\, k\}$.
\end{enumerate}
Summing up,
\begin{fact}The critical points of\,  $||\mu||^{2}$ are $K$-orbits
through the maximally entangled states of rank $1,\ldots,\, N$, i.e.\ states of
the form
\begin{gather*}
v_{k}=\frac{1}{\sqrt{k}}\sum_{i=1}^{k}\ket{i,i},
\end{gather*}
where $k\in\{1,\ldots,N\}$. By the formula (\ref{eq:var3}) the states
belonging to each orbit $K.v_{k}$ are those with the maximal variance
$\mathrm{Var}([v])$ among all states in $G.v_{k}$, i.e.\ in the SLOCC class
of state $v_{k}$ .
\end{fact}

Once we know the critical sets of $||\mu||^{2}$ we can easily compute
the distance of each SLOCC class polytope $\Psi(\overline{G.[v_{k}]})$
from the origin, i.e.\ the numbers $d([v_{k}])$.

\begin{fact}The distance of the SLOCC class polytope $\Psi(\overline{G.[v_{k}]})$
from the origin is given by
\begin{gather*}
d([v_{k}])=\sqrt{\mathrm{tr}\left(\rho_{1}([v_{k}])-\frac{1}{N}I\right)^{2}
+\mathrm{tr}\left(\rho_{2}([v_{k}])-\frac{1}{N}I\right)^{2}}=\\
=\frac{\sqrt{2\left(k(N-k)^{2}+k^{2}(N-k)\right)}}{Nk}
\end{gather*}
where $k\in\{1,\ldots,\, N\}$. \end{fact} The next task is to calculate the
Morse index for each critical set. Let us recall that by
Section~\ref{sec:The-function-mu2} it reduces to
calculation of $\mathrm{Hess}\mu_{\mu^{\ast}[v_{k}]}$ restricted to
$\left(T_{[v_{k}]}G.[v_{k}]\right)^{\bot\omega}$. First we notice that
\begin{gather*}
\left(T_{[v_{k}]}G.[v_{k}]\right)^{\bot\omega}=
\mbox{Span}\left\{ \ket{m,n},\, i\ket{m,n}:\, m,n\in\{k+1,\ldots,N\}\right\} .
\end{gather*}
It is thus enough to consider a perturbed state of the form
$v=v_{k}+\sum_{i,j=k+1}^{N}C_{ij}\ket{i,j}$ and the function
\begin{gather}
\mu_{\mu^{\ast}}([v])=\frac{\bk v{\mu^{\ast}([v_{k}])|v}}{\bk vv}=
\frac{\bk v{\rho_{1}([v_{k}])\otimes I+I\otimes\rho([v_{k}])|v}}{\bk vv}.
\label{eq:2particles}
\end{gather}
Making use of the assumption that $a_{k+1},\ldots,a_{N}=0$ for $v_{k}$
we can rewrite (\ref{eq:2particles}) in the form
\begin{gather}
\mu_{\mu^{\ast}}([v])=\frac{\bra{v_{k}}\mu([v_{k}])\ket{v_{k}}}{\bk vv}-
\frac{2}{N}\frac{\sum_{ij=k+1}^{N}|C_{ij}|^{2}}{\bk vv},
\label{eq:index dist}
\end{gather}
Hence any perturbation of $\mu_{\mu^{\ast}}$ in a direction from
$\left(T_{[v_{k}]}G.[v_{k}]\right)^{\bot\omega}$ decreases its value. Finally
thus,
\begin{fact}The Morse index at critical set $K.[v_{k}]$ is given
by
\begin{gather*}
\mathrm{ind}([v_{k}])=2\left(N-k\right){}^{2},
\end{gather*}
where $k\in\{1,\ldots,\, N\}$.
\end{fact}

\subsection{Three qubits}

The simplest non-bipartite case involves three-qubits. The Hilbert space of the
system is
$\mathcal{H}=\mathbb{C}^{2}\otimes\mathbb{C}^{2}\otimes\mathbb{C}^{2}$. The
SLOCC operations are given by the group $G=SL(2,\mathbb{C})^{\times3}$ and the
local unitary operations by the group $K=SU(2)^{\times3}$. We know that in
order to find SLOCC classes we have to find eigenvectors of the equation
(\ref{eig}) which for three qubit reads
\begin{gather*}
\left(\left(\rho_{1}-\frac{1}{2}\right)\otimes I\otimes I+I\otimes\left(\rho_{2}-
\frac{1}{2}\right)\otimes I+I\otimes I\otimes\left(\rho_{3}-
\frac{1}{2}\right)\right)v=\lambda v.
\end{gather*}
Following the algorithm given in Section~\ref{sec:The-SLOCC-invariant} we have
to find all states for which all matrices $\rho_{i}$ are maximally mixed. To
this end we use the canonical form of the state which by Theorem \ref{thm1} is
given by
\begin{equation}
v=p\ket{011}+q\ket{101}+r\ket{110}+s\ket{111}+z\ket{000},\label{Acin}
\end{equation}
where $z\in\mathbb{C}$ and other coefficients are real and positive. The
reduced one-qubit density matrices of state $v$ in the form (\ref{Acin}) can be
easily calculated:
\begin{gather*}
\rho_{1}([v])=\left(\begin{array}{cc}
|z|^{2}+p^{2} & ps\\
ps & q^{2}+r^{2}+s^{2}
\end{array}\right),\\
\rho_{2}([v])=\left(\begin{array}{cc}
|z|^{2}+q^{2} & qs\\
qs & p^{2}+r^{2}+s^{2}
\end{array}\right),\\
\rho_{3}([v])=\left(\begin{array}{cc}
|z|^{2}+r^{2} & rs\\
rs & p^{2}+q^{2}+s^{2}
\end{array}\right).
\end{gather*}
We require that all diagonal elements of $\rho_{i}$ are the same
and the off-diagonal vanish. It leads to the equations
\begin{gather*}
p^{2}=q^{2}=r^{2}=|z|^{2}-s^{2},\quad ps=qs=rs=0,
\end{gather*}
having two solutions. The first one, $s=0$ and $p^{2}=q^{2}=r^{2}=|z|^{2}$,
corresponds to the state
\begin{gather*}
v_{1}=\frac{1}{2}\left(\ket{011}+\ket{101}+\ket{110}+\ket{000}\right),
\end{gather*}
and the second one is $s\neq0$ and $p=q=r=0$, i.e.
\begin{gather*}
v_{2}=\frac{1}{\sqrt{2}}\left(\ket{000}+\ket{111}\right),
\end{gather*}
where in both cases we have already removed the freedom of phase stemming
from $z$. Notice, however, that the states $v_{1}$ and $v_{2}$
are local unitary equivalent%
\footnote{For yet another justification of local unitary equivalence see \cite{SK11},
where it is shown that $K.v_{2}$ is Lagrangian.%
}, i.e.
\begin{gather*}
U\otimes U\otimes U.v_{2}=v_{1},
\end{gather*}
where
\begin{gather*}
U=\frac{1}{\sqrt{2}}\left(\begin{array}{cc}
1 & 1\\
1 & -1
\end{array}\right).
\end{gather*}
In this way we find that the SLOCC class corresponding to $\mu^{-1}(0)/K$ can
be represented by a single state
$v_{GHZ}=\frac{1}{\sqrt{2}}\left(\ket{000}+\ket{111}\right)$ which is the
well-known Greenberger-Horne-Zeilinger (GHZ) state. What is left is to find
remaining SLOCC classes in the null cone $N$. To this end it is enough to
consider diagonal $\rho_{i}$ and we assume that
\begin{gather*}
\rho_{i}=\left(\begin{array}{cc}
1-p_{i} & 0\\
0 & p_{i}
\end{array}\right),
\end{gather*}
where $0\leq p_{i}\leq\frac{1}{2}$. This, of course, means that
\begin{gather*}
\rho_{i}-\frac{1}{2}I=\left(\begin{array}{cc}
\frac{1}{2}-p_{i} & 0\\
0 & p_{i}-\frac{1}{2}
\end{array}\right)=\left(\begin{array}{cc}
\lambda_{i} & 0\\
0 & -\lambda_{i}
\end{array}\right),
\end{gather*}
where $\frac{1}{2}\geq\lambda_{i}\geq0$, i.e.\ $\mu^{\ast}([v])$ belongs to the
positive Weyl chamber. We can always write $\mu^{\ast}([v])$ as the $8\times8$
diagonal matrix in the basis $\{\ket{ijk}\}$
\begin{gather}
\mu^{\ast}([v])=\mbox{diag}\left(f_{1},\, f_{2},\, f_{3},\,
f_{4},-f_{4},-f_{3},-f_{2},-f_{1}\right),
\label{eq:muspectrum}
\end{gather}
where
\begin{gather}
f_{1}=\lambda_{1}+\lambda_{2}+\lambda_{3},\,\, f_{2}=\lambda_{1}+\lambda_{2}-
\lambda_{3},\,\, f_{3}=\lambda_{1}-\lambda_{2}+\lambda_{3},\nonumber \\
f_{4}=\lambda_{1}-\lambda_{2}-\lambda_{3}=-f_{1}+f_{2}+f_{3}.
\label{eq:spectrum}
\end{gather}
We are interested in the structure of spectrum of $\mu^{\ast}([v])$ on the
Kirwan polytope which by Theorem~\ref{polytope} is given by the inequalities
\begin{gather}
\frac{1}{2}\geq-\lambda_{1}+\lambda_{2}+\lambda_{3},\,\,\frac{1}{2}\geq\lambda_{1}
-\lambda_{2}+\lambda_{3},\,\,\frac{1}{2}\geq\lambda_{1}+\lambda_{2}-\lambda_{3},
\label{eq:ineq}
\end{gather}
where $\frac{1}{2}\geq\lambda_{i}\geq0$. In order to find the SLOCC classes in
the null cone $N$ we assume that $\mu^{\ast}([v])\neq0$ and analyze eigenspaces
of $\mu^{\ast}([v])$ on the Kirwan polytope. The following fact is a
straightforward consequence of (\ref{eq:muspectrum}) and (\ref{eq:spectrum}),

\begin{fact}Assume that $\mu^{\ast}([v])\neq0$, then $\mu^{\ast}([v])$
has no eigenspaces of dimension $5$, $6$, $7$ and $8$. \end{fact}

Therefore we have to consider only a situation when the spectrum is
non-degenerate or $2$, $3$, $4$-fold degenerate. Notice, that if the spectrum
of $\mu^{\ast}([v])$ is multiplicity free then eigenvectors are separable
vectors $\ket{ijk}$. However only the vector $\ket{000}$ gives $\mu(\ket{000})$
in the positive Weyl chamber. Hence $\ket{000}$ represents the first SLOCC
class contained in the null cone. Assume now that two diagonal entries of
(\ref{eq:muspectrum}) are the same. There are \textit{a priori} ${8 \choose
2}=28$ possibilities to consider. However all of them lead to states for which
either reduced density matrices are not diagonal or for which degeneracies in
the spectrum of $\mu^{\ast}([v])$ are of higher order. Let us now consider the
case when spectrum of $\mu^{\ast}([v])$ has a 3-fold degenerate eigenvalue.
Notice, however, that the equality of any three among four $f_{i}$-s implies
the equality of all four $f_{i}$. This means that 3-fold degenerate eigenvalues
come from choosing two $f_{i}$-s and one $-f_{k}$. Going over these
possibilities we find that the only nontrivial one is given by
$f_{2}=f_{3}=-f_{4}$, i.e.\ $\lambda_{1}=\lambda_{2}=\lambda_{3}=\alpha$ and
$\mu^{\ast}([v])=\{3\alpha,\,\alpha,\,\alpha,\,-\alpha,\,\alpha,\,-\alpha,\,-\alpha,\,-3\alpha\}$.
The general state from the eigenspace corresponding to the eigenvalue $\alpha$
is hence
\begin{gather*}
v=z_{1}\ket{001}+z_{2}\ket{010}+z_{3}\ket{100},
\end{gather*}
the matrices $\rho_{i}-\frac{1}{2}I$ for $[v]$ read
\begin{gather*}
\left(\begin{array}{cc}
|z_{1}|^{2}+|z_{2}|^{2}-\frac{1}{2} & 0\\
0 & |z_{3}|^{2}-\frac{1}{2}
\end{array}\right),\,\left(\begin{array}{cc}
|z_{1}|^{2}+|z_{3}|^{2}-\frac{1}{2} & 0\\
0 & |z_{2}|^{2}-\frac{1}{2}
\end{array}\right),\\
\,\left(\begin{array}{cc}
|z_{2}|^{2}+|z_{3}|^{2}-\frac{1}{2} & 0\\
0 & |z_{1}|^{2}-\frac{1}{2}
\end{array}\right).
\end{gather*}
The solution satisfying $\lambda_{1}=\lambda_{2}=\lambda_{3}$ is
$|z_{1}|^{2}=|z_{2}|^{2}=|z_{3}|^{2}$, so
\begin{gather*}
v=\frac{1}{\sqrt{3}}\left(\ket{001}+\ket{010}+\ket{100}\right).
\end{gather*}
Notice also that the state corresponding to the $-\alpha$-eigenspace is LU
equivalent to $v$ and is not mapped to the Kirwan polytope. Finally for
$4$-fold degenerate eigenvalues we have the following possibilities
\begin{enumerate}
\item $f_{1}=f_{2}=f_{3}=f_{4}$, in this case
\begin{gather*}
\mu^{\ast}([v])=\{\lambda_{1},\,\lambda_{1},\,\lambda_{1},\,\lambda_{1},\,
-\lambda_{1},\,-\lambda_{1},\,-\lambda_{1},\,-\lambda_{1}\},
\end{gather*}
i.e.\ $\lambda_{2}=\lambda_{3}=0$. It is easy to see that this corresponds
to $v=\ket 0\otimes\left(\ket{00}+\ket{11}\right)$. At the same time
we get that the state corresponding to the eigenvalue $-\lambda_{1}$
is $\ket 1\otimes\left(\ket{00}+\ket{11}\right)$ which is LU equivalent
to $v$.
\item $f_{1}=f_{2}=-f_{3}=-f_{4}$, in this case
\begin{gather*}
\mu^{\ast}([v])=\{\lambda_{2},\,\lambda_{2},\,-\lambda_{2},\,-
\lambda_{2},\,\lambda_{2},\,\lambda_{2},\,-\lambda_{2},\,-\lambda_{2}\}
\end{gather*}
i.e.\ $\lambda_{1}=\lambda_{3}=0$ which corresponds to $v=\ket{000}+\ket{101}$.
The state corresponding to the eigenvalue $-\lambda_{2}$ is $\ket{010}+\ket{111}$
which is LU equivalent to $v$.
\item $f_{1}=-f_{2}=f_{3}=-f_{4}$, in this case
\begin{gather*}
\mu^{\ast}([v])=\{\lambda_{3},\,-\lambda_{3},\,\lambda_{3},\,-
\lambda_{3},\,\lambda_{3},\,-\lambda_{3},\,\lambda_{3},\,-\lambda_{3}\}
\end{gather*}
i.e.\ $\lambda_{1}=\lambda_{2}=0$. This corresponds to $v=\ket{000}+\ket{110}$.
The state corresponding to the eigenvalue $-\lambda_{3}$ is $\ket{001}+\ket{111}$
is LU equivalent to $v$.
\item $f_{1}=f_{2}=0$, i.e.\ $\lambda_{2}=-\lambda_{1}$ and $\lambda_{3}=0$.
Now,
\begin{gather*}
\mu^{\ast}([v])=\{0,\,0,\,2\lambda_{1},\,2\lambda_{1},\,-2\lambda_{1},\,-2\lambda_{1},\,0,\,0\}.
\end{gather*}
This, however, by positivity of $\lambda_{i}$ implies that also $\lambda_{1}=0$.
Similarly we exclude cases when other pairs of $f_{i}$-s are zero.
\end{enumerate}
Summing up we obtained exactly six SLOCC classes of states which are
given by the $G$-orbits through \cite{DWC00}
\begin{gather*}
v_{GHZ}=\frac{1}{\sqrt{2}}\left(\ket{000}+\ket{111}\right),\\
v_{W}=\frac{1}{\sqrt{3}}\left(\ket{100}+\ket{010}+\ket{001}\right),\\
v_{B1}=\frac{1}{\sqrt{2}}\left(\ket{000}+\ket{011}\right),\,\, v_{B2}=
\frac{1}{\sqrt{2}}\left(\ket{000}+\ket{101}\right),\\
v_{B3}=\frac{1}{\sqrt{2}}\left(\ket{000}+\ket{110}\right),\,\, v_{SEP}=
\ket{000}.
\end{gather*}
\begin{fact}The critical points of $||\mu||^{2}$ for three qubits
are $K$-orbits through states $v_{i}$. By formula (\ref{eq:var3})
states belonging to each orbit $K.v_{k}$ are those with the maximal
variance $\mathrm{Var}([v])$ among all states in $G.v_{k}$, i.e.\ in
the SLOCC class of the state $v_{k}$ . \end{fact}

We can now easily compute the distance of each SLOCC class polytope
$\Psi(\overline{G.[v_{k}]})$ from the origin, i.e.\ the number $d([v_{k}])$.

\begin{fact}The distance of the SLOCC class polytope $\Psi(\overline{G.[v_{k}]})$
from the origin is given by
\begin{gather*}
d([v_{k}])=\sqrt{\sum_{i=1}^{3}\mathrm{tr}\left(\rho_{i}([v_{k}])-\frac{1}{2}I\right)^{2}}=
\begin{cases}
0 & \mathrm{for\ }v_{GHZ},\\
\sqrt{\frac{1}{6}} & \mathrm{for\ }v_{W},\\
\sqrt{\frac{1}{2}} & \mathrm{for\ }v_{B1},\, v_{B2},\, v_{B3}\\
\sqrt{\frac{3}{2}} & \mathrm{for\ }v_{SEP}.
\end{cases}
\end{gather*}
\end{fact}
To calculate the Morse index for each critical set $K.v_{k}$
we have to find $\mathrm{Hess}\mu_{\mu^{\ast}}$ restricted to $\left(T_{v_{k}}G.[v_{k}]\right)^{\bot\omega}$.
Let us first notice that $\left(T_{v_{GHZ}}G.[v_{GHZ}]\right)^{\bot\omega}=0$
as the $G$-orbit through the state $v_{GHZ}$ is dense in $\mathbb{P}(\mathcal{H})$
(see for example \cite{SWK12}). This means that $\mathrm{ind}(v_{GHZ})=0$.
In case of $v_{W}$ one has
\begin{gather*}
\left(T_{v_{W}}G.[v_{W}]\right)^{\bot\omega}=\mbox{Span}\left\{ \ket{111},\, i\ket{111}\right\} .
\end{gather*}
We can hence consider a perturbed state in the form $v=v_{W}+\epsilon_{1}\ket{111}+i\epsilon_{2}\ket{111}$.
Simple calculations give
\begin{gather*}
\mu_{\mu^{\ast}}([v])=\frac{\bk{v_{W}}{\mu^{\ast}([v_{W}])|v_{W}}}{1+\epsilon_{1}^{2}+\epsilon_{1}^{2}}-\frac{(\epsilon_{1}^{2}+\epsilon_{2}^{2})}{2(1+\epsilon_{1}^{2}+\epsilon_{1}^{2})}.
\end{gather*}
It is now clear that $\mathrm{ind}(v_{W})=2$. Similar calculations
show that for bi-separable states the Morse index is $\mathrm{ind}(v_{B_{k}})=6$
and for the separable state $\mathrm{ind}(v_{SEP})=8$.

\subsection{Four qubits}

In \cite{VDDV02} it was shown that there are nine families of inequivalent
SLOCC classes of four qubits. In this subsection we show how to calculate the
main group of families of SLOCC classes, i.e. $\mu^{-1}(0)/K$ and prove that among other seven families calculated in \cite{VDDV02} there are five
which are closure equivalent to it.

\paragraph{The main family of SLOCC classes, i.e. $\mu^{-1}(0)/K$}

Applying Theorem \ref{thm1} to four-qubit states one gets the normal
form:
\begin{gather*}
v=z_{1}\ket{0000}+z_{2}\ket{0011}+z_{3}\ket{0101}+z_{4}\ket{0110}+a_{5}\ket{0111}+z_{6}\ket{1001}+z_{7}\ket{1010}+\\
+a_{8}\ket{1011}+z_{9}\ket{1100}+a_{10}\ket{1101}+a_{11}\ket{1110}+a_{12}\ket{1111},
\end{gather*}
where $z_{i}$ are complex and $a_{i}$ are real positive. The one-qubit
reduced density matrices are given by
\begin{gather*}
\rho_{1}=\left(\begin{array}{cc}
|z_{1}|^{2}+|z_{2}|^{2}+|z_{3}|^{2}+|z_{4}|^{2}+a_{5}^{2} & \bar{z}_{2}a_{8}+\bar{z}_{3}a_{10}+\bar{z}_{4}a_{11}+a_{5}a_{12}\\
z_{2}a_{8}+z_{3}a_{10}+z_{4}a_{11}+a_{5}a_{12} & |z_{6}|^{2}+|z_{7}|^{2}+a_{8}^{2}+|z_{9}|^{2}+a_{10}^{2}+a_{11}^{2}+a_{12}^{2}
\end{array}\right),\\
\rho_{2}=\left(\begin{array}{cc}
|z_{1}|^{2}+|z_{2}|^{2}+|z_{6}|^{2}+|z_{7}|^{2}+a_{8}^{2} & \bar{z}_{2}a_{5}+\bar{z}_{6}a_{10}+\bar{z}_{7}a_{11}+a_{8}a_{12}\\
z_{2}a_{5}+z_{6}a_{10}+z_{7}a_{11}+a_{8}a_{12} & |z_{3}|^{2}+|z_{4}|^{2}+a_{5}^{2}+|z_{9}|^{2}+a_{10}^{2}+a_{11}^{2}+a_{12}^{2}
\end{array}\right),\\
\rho_{3}=\left(\begin{array}{cc}
|z_{1}|^{2}+|z_{3}|^{2}+|z_{6}|^{2}+|z_{9}|^{2}+a_{10}^{2} & \bar{z}_{3}a_{5}+\bar{z}_{6}a_{8}+\bar{z}_{9}a_{11}+a_{10}a_{12}\\
z_{3}a_{5}+z_{6}a_{8}+z_{9}a_{11}+a_{10}a_{12} & |z_{2}|^{2}+|z_{4}|^{2}+a_{5}^{2}+|z_{7}|^{2}+a_{8}^{2}+a_{11}^{2}+a_{12}^{2}
\end{array}\right),\\
\rho_{3}=\left(\begin{array}{cc}
|z_{1}|^{2}+|z_{4}|^{2}+|z_{7}|^{2}+|z_{9}|^{2}+a_{11}^{2} & \bar{z}_{4}a_{5}+\bar{z}_{7}a_{8}+\bar{z}_{9}a_{10}+a_{11}a_{12}\\
z_{4}a_{5}+z_{7}a_{8}+z_{9}a_{10}+a_{11}a_{12} & |z_{2}|^{2}+|z_{3}|^{2}+a_{5}^{2}+|z_{6}|^{2}+a_{8}^{2}+a_{10}^{2}+a_{12}^{2}
\end{array}\right),
\end{gather*}
We require all diagonal elements of $\rho_{i}$ to be equal to each
other and the off-diagonal to be zero. The equations for the diagonal
elements give:
\begin{gather}
a_{5}^{2}-a_{8}^{2}=|z_{6}|^{2}+|z_{7}|^{2}-|z_{3}|^{2}-|z_{4}|^{2}\nonumber \\
a_{8}^{2}-a_{10}^{2}=|z_{3}|^{2}+|z_{9}|^{2}-|z_{2}|^{2}-|z_{7}|^{2}\nonumber \\
a_{8}^{2}+a_{10}^{2}=|z_{1}|^{2}+|z_{4}|^{2}-|z_{6}|^{2}-a_{12}^{2}\nonumber \\
|z_{4}|^{2}+|z_{7}|^{2}=|z_{3}|^{2}+|z_{6}|^{2}\label{eq:diag-con}\\
a_{10}^{2}=a_{11}^{2}\nonumber
\end{gather}
The equations for the off-diagonal elements can be written as a matrix
equation $Av=0$, where
\begin{gather*}
A=\left(\begin{array}{cccc}
a_{12} & z_{2} & z_{3} & z_{4}\\
z_{2} & a_{12} & z_{6} & z_{7}\\
z_{3} & z_{6} & a_{12} & z_{9}\\
z_{4} & z_{7} & z_{9} & a_{12}
\end{array}\right),\,\,\, v=\left(\begin{array}{c}
a_{5}\\
a_{8}\\
a_{10}\\
a_{11}
\end{array}\right).
\end{gather*}
Notice that parameters in the matrix $A$ and the vector $v$ are disjoint and
$A$ is a symmetric complex matrix. The solution $a_{10},\, a_{11},\, a_{5},\,
a_{8}=0$ is always a legitimate one. The other possibility would be
$\mbox{det}A=0$ and $v\in\mbox{Ker}(A)$. It can be checked by straightforward
although tedious calculations that this possibility does not give any new
solutions. Hence the only solution is $a_{10},\, a_{11},\, a_{5},\, a_{8}=0$
and the equations (\ref{eq:diag-con}) reduce to
\begin{gather}
0=|z_{6}|^{2}+|z_{7}|^{2}-|z_{3}|^{2}-|z_{4}|^{2}\nonumber \\
0=|z_{3}|^{2}+|z_{9}|^{2}-|z_{2}|^{2}-|z_{7}|^{2}\nonumber \\
0=|z_{1}|^{2}+|z_{4}|^{2}-|z_{6}|^{2}-a_{12}^{2}\nonumber \\
0=|z_{3}|^{2}+|z_{6}|^{2}-|z_{4}|^{2}-|z_{7}|^{2}\label{eq:diag-con-1}
\end{gather}
These are four equations for $8$ unknowns. Writing them in the matrix
form we easily see that solutions are given by
\begin{gather*}
|z_{3}|=|z_{7}|,\,\,|z_{4}|=|z_{6}|,\,\,|z_{2}|=|z_{9}|,\,\,|z_{1}|=a_{12}
\end{gather*}
Hence, up to normalization the states from $\mu^{-1}(0)/K$ are given by
\begin{gather*}
v=\alpha_{1}(\ket{0000}+\ket{1111})+\alpha_{2}(\ket{0011}+\ket{1100})+\\
\alpha_{3}(\ket{0101}+\ket{1010})+\alpha_{4}(\ket{0110}+\ket{1001}).
\end{gather*}
It is also clear that on the projective level the reduced space $\mu^{-1}(0)/K$
is a complex projective space. Summing up
\begin{gather*}
\mu^{-1}(0)/K=\mathbb{P}(\mathcal{H}_{0}),
\end{gather*}
where
\begin{gather*}
\mathcal{H}_{0}=\mbox{Span}_{\mathbb{C}}\{v_{1},\, v_{2},\, v_{3},\, v_{4}\},
\end{gather*}
and
\begin{gather*}
v_{1}=\ket{0000}+\ket{1111},\,\, v_{2}=\ket{0011}+\ket{1100},\\
v_{3}=\ket{0101}+\ket{1010},\,\, v_{4}=\ket{0110}+\ket{1001}.
\end{gather*}
This is the class denoted in \cite{VDDV02} by $G_{abcd}$. Notice also that
when all coefficients $\alpha_{i}\neq0$ then the state is stable, i.e. $\dim
G.[v]=\dim G$. It means that families of SLOCC classes associated to these
states are given by single $G$-orbits. We will now show that families
$L_{abc_{2}}$, $L_{a_{2}b_{2}}$, $L_{ab_{3}}$, $L_{a_{4}}$,
$L_{a_{2}0_{3\oplus\bar{1}}}$ from \cite{VDDV02} are closure equivalent to
$G_{abcd}$. The general strategy for each family is to divide the state into
linear combination of the state belonging to $G_{abcd}$ which we denote by
$\ket v$ and the rest which we denote by $\ket w$. In all cases the
stabilizer of $\ket v$ in $G$ contains some subgroup $P(\ket{v})$ of the
complex torus $T^\mathbb{C}$. Acting with the certain one-parameter subgroup
$P_\alpha \subset P(\ket{v})$ on the whole state asymptotically we can get
rid of $\ket w$.

\subsubsection*{Family $L_{abc_{2}}$}
\begin{gather*}
\ket{L_{abc_{2}}}=\ket v+\ket w,\,\,\,\,\,\,\ket w=\ket{0110}\,,\\
\ket v=\frac{a+b}{2}\left(\ket{0000}+\ket{1111}\right)+\frac{a-b}{2}\left(\ket{0011}+\ket{1100}\right)+c\left(\ket{0101}+\ket{1010}\right)\,.
\end{gather*}
\[
P \left(\ket v\right)=\left\{ \left(\begin{array}{cc}
e^{\alpha} & 0\\
0 & e^{-\alpha}
\end{array}\right)\otimes\left(\begin{array}{cc}
e^{\beta} & 0\\
0 & e^{-\beta}
\end{array}\right)\otimes\left(\begin{array}{cc}
e^{\gamma} & 0\\
0 & e^{-\gamma}
\end{array}\right)\otimes\left(\begin{array}{cc}
e^{\delta} & 0\\
0 & e^{-\delta}
\end{array}\right)\left|\alpha=-\beta=-\gamma=\delta\right.\right\} .
\]
Choosing the one-parameter subgroup $P_{\alpha}\subset P\left(\ket
v\right)$
\[
P_{\alpha}=\left\{\left(\begin{array}{cc}
e^{\alpha} & 0\\
0 & e^{-\alpha}
\end{array}\right)\otimes\left(\begin{array}{cc}
e^{-\alpha} & 0\\
0 & e^{\alpha}
\end{array}\right)\otimes\left(\begin{array}{cc}
e^{-\alpha} & 0\\
0 & e^{\alpha}
\end{array}\right)\otimes\left(\begin{array}{cc}
e^{\alpha} & 0\\
0 & e^{-\alpha}
\end{array}\right),\,\alpha\in\mathbb{R}\right\}\,.
\]
we obtain
\[
P_{\alpha}\ket v=\ket v,\,\mathrm{\lim_{\alpha\rightarrow\infty}}P_{\alpha}\ket w=\mathrm{\lim_{\alpha\rightarrow\infty}e^{-4\alpha}}\ket w=0\,.
\]
And hence $\lim_{\alpha\rightarrow\infty}P_{\alpha}\ket{L_{abc_{2}}}\in G_{abcd}$.

\subsubsection*{Family $L_{a_{2}b_{2}}$}
\begin{gather*}
\ket{L_{a_{2}b_{2}}}=\ket v+\ket w,\,\,\,\ket w=\ket{0011}+\ket{1100}\,.\\
\ket v=a\left(\ket{0000}+\ket{1111}\right)+b\left(\ket{0101}+\ket{1010}\right)\\
\mathrm{P\left(\ket v\right)}=\left\{ \left(\begin{array}{cc}
e^{\alpha} & 0\\
0 & e^{-\alpha}
\end{array}\right)\otimes\left(\begin{array}{cc}
e^{\beta} & 0\\
0 & e^{-\beta}
\end{array}\right)\otimes\left(\begin{array}{cc}
e^{\gamma} & 0\\
0 & e^{-\gamma}
\end{array}\right)\otimes\left(\begin{array}{cc}
e^{\delta} & 0\\
0 & e^{-\delta}
\end{array}\right)\left|\alpha=-\gamma\,,\beta=-\delta\right.\right\} .
\end{gather*}
By choosing the one-parameter subgroup
$P_{\alpha}\subset P \left(\ket v\right)$
\[
P_{\alpha}=\left\{\left(\begin{array}{cc}
1 & 0\\
0 & 1
\end{array}\right)\otimes\left(\begin{array}{cc}
e^{\alpha} & 0\\
0 & e^{-\alpha}
\end{array}\right)\otimes\left(\begin{array}{cc}
1 & 0\\
0 & 1
\end{array}\right)\otimes\left(\begin{array}{cc}
e^{-\alpha} & 0\\
0 & e^{\alpha}
\end{array}\right),\,\alpha\in\mathbb{R}\right\}\,.
\]
we get
\[
P_{\alpha}\ket v=\ket v,\,\mathrm{\lim_{\alpha\rightarrow\infty}}P_{\alpha}\ket w=\mathrm{\lim_{\alpha\rightarrow\infty}e^{-2\alpha}}\ket w=0\,.
\]
Consequently,
$\lim_{\alpha\rightarrow\infty}P_{\alpha}\ket{L_{a_{2}b_{2}}}\in G_{abcd}$.

\subsubsection*{Family $L_{ab_{3}}$}
\begin{gather*}
\ket{L_{ab_{3}}}=\ket v+\ket w,\\
\ket v=a\left(\ket{0000}+\ket{1111}\right)+\frac{a+b}{2}\left(\ket{0101}+\ket{1010}\right)+\frac{a-b}{2}\left(\ket{0110}+\ket{1001}\,\right),\\
\ket w=\frac{i}{\sqrt{2}}\left(\ket{0001}+\ket{0010}+\ket{0111}+\ket{1011}\right),\\
P \left(\ket v\right)=\left\{ \left(\begin{array}{cc}
e^{\alpha} & 0\\
0 & e^{-\alpha}
\end{array}\right)\otimes\left(\begin{array}{cc}
e^{\beta} & 0\\
0 & e^{-\beta}
\end{array}\right)\otimes\left(\begin{array}{cc}
e^{\gamma} & 0\\
0 & e^{-\gamma}
\end{array}\right)\otimes\left(\begin{array}{cc}
e^{\delta} & 0\\
0 & e^{-\delta}
\end{array}\right)\left|\alpha=\beta=-\gamma=-\delta\right.\right\} .
\end{gather*}

Acting with the one-parameter subgroup
$P_{\alpha}\subset P\left(\ket v\right)$
\[
P_{\alpha}=\left\{\left(\begin{array}{cc}
e^{\alpha} & 0\\
0 & e^{-\alpha}
\end{array}\right)\otimes\left(\begin{array}{cc}
e^{\alpha} & 0\\
0 & e^{-\alpha}
\end{array}\right)\otimes\left(\begin{array}{cc}
e^{-\alpha} & 0\\
0 & e^{\alpha}
\end{array}\right)\otimes\left(\begin{array}{cc}
e^{-\alpha} & 0\\
0 & e^{\alpha}
\end{array}\right),\,\alpha\in\mathbb{R}\right\}\,.
\]
we produce
\[
P_{\alpha}\ket v=\ket v,\,\mathrm{\lim_{\alpha\rightarrow\infty}}P_{\alpha}\ket w=\mathrm{\lim_{\alpha\rightarrow\infty}e^{-2\alpha}}\ket w=0\,.
\]
Thus, again $\lim_{\alpha\rightarrow\infty}P_{\alpha}\ket{L_{ab_{3}}}\in
G_{abcd}$.

\paragraph{Family $L_{a_{4}}$}

\begin{gather*}
\ket{L_{a{}_{4}}}=\ket v+\ket w\\
\ket v=a\left(\ket{0000}+\ket{0101}+\ket{1010}+\ket{1111}\,\right),\\
\ket w=i\ket{0001}+\ket{0110}-i\ket{1011}.\\
P \left(\ket v\right)=\left\{ \left(\begin{array}{cc}
e^{\alpha} & 0\\
0 & e^{-\alpha}
\end{array}\right)\otimes\left(\begin{array}{cc}
e^{\beta} & 0\\
0 & e^{-\beta}
\end{array}\right)\otimes\left(\begin{array}{cc}
e^{\gamma} & 0\\
0 & e^{-\gamma}
\end{array}\right)\otimes\left(\begin{array}{cc}
e^{\delta} & 0\\
0 & e^{-\delta}
\end{array}\right)\left|\alpha=-\gamma\,,\beta=-\delta\right.\right\} .
\end{gather*}
 Here we choose the one-parameter subgroup $P_{\alpha}\subset P \left(\ket v\right)$
\[
P_{\alpha}=\left\{\left(\begin{array}{cc}
e^{2\alpha} & 0\\
0 & e^{-2\alpha}
\end{array}\right)\otimes\left(\begin{array}{cc}
e^{\alpha} & 0\\
0 & e^{-\alpha}
\end{array}\right)\otimes\left(\begin{array}{cc}
e^{-2\alpha} & 0\\
0 & e^{2\alpha}
\end{array}\right)\otimes\left(\begin{array}{cc}
e^{-\alpha} & 0\\
0 & e^{\alpha}
\end{array}\right),\,\alpha\in\mathbb{R} \right\}\,.
\]
to obtain
\[
P_{\alpha}\ket v=\ket v,\,\mathrm{\lim_{\alpha\rightarrow\infty}}P_{\alpha}\ket w=\mathrm{\lim_{\alpha\rightarrow\infty}e^{-2\alpha}}\ket w=0\,.
\]
and conclude that $\lim_{\alpha\rightarrow\infty}P_{\alpha}\ket{L_{a_{4}}}\in
G_{abcd}$.

\subsubsection*{Family $L_{a_{2}0_{3\oplus\bar{1}}}$}
\begin{gather*}
\ket{L_{a_{2}0_{3\oplus\bar{1}}}}=\ket v+\ket w,\\
\ket v=a\left(\ket{0000}+\ket{1111}\,\right),\\
\ket w=\ket{0011}+\ket{0101}+\ket{0110}\,.\\
P\left(\ket v\right)=\left\{ \left(\begin{array}{cc}
e^{\alpha} & 0\\
0 & e^{-\alpha}
\end{array}\right)\otimes\left(\begin{array}{cc}
e^{\beta} & 0\\
0 & e^{-\beta}
\end{array}\right)\otimes\left(\begin{array}{cc}
e^{\gamma} & 0\\
0 & e^{-\gamma}
\end{array}\right)\otimes\left(\begin{array}{cc}
e^{\delta} & 0\\
0 & e^{-\delta}
\end{array}\right)\left|\delta=-\alpha-\beta-\gamma\right.\right\} .
\end{gather*}

By choosing the one-parameter subgroup $P_{\alpha}\subset P\left(\ket
v\right)$

\[
P_{\alpha}=\left\{\left(\begin{array}{cc}
e^{3\alpha} & 0\\
0 & e^{-3\alpha}
\end{array}\right)\otimes\left(\begin{array}{cc}
e^{-\alpha} & 0\\
0 & e^{+\alpha}
\end{array}\right)\otimes\left(\begin{array}{cc}
e^{-\alpha} & 0\\
0 & e^{\alpha}
\end{array}\right)\otimes\left(\begin{array}{cc}
e^{-\alpha} & 0\\
0 & e^{\alpha}
\end{array}\right),\,\alpha\in\mathbb{R}\right\}\,.
\]

\[
P_{\alpha}\ket v=\ket v,\,\mathrm{\lim_{\alpha\rightarrow\infty}}P_{\alpha}\ket w=\mathrm{\lim_{\alpha\rightarrow\infty}e^{-4\alpha}}\ket w=0\,.
\]

And hence $\lim_{\alpha\rightarrow\infty}P_{\alpha}\ket{L_{a_{2}0_{3\oplus\bar{1}}}}\in G_{abcd}$.

\section{Indistinguishable particles}

\label{sec:indistinguishable} After calculating examples for distinguishable
particles we switch to the indistinguishable case. In the following we show
in particular that for the system of arbitrary many
two-state bosons the groups of families of SLOCC classes in the null cone
always contain only one critical set of the total variance of state.
Moreover, for two $N$-state bosons and fermions, similarly to bipartite and three qubits cases, all SLOCC classes
are detected by the total variance of state.

\subsection{The momentum map for bosonic and fermionic particles}

Let $K=SU(N)$ act by the diagonal action on the Hilbert space of $L$ bosons
($\mathcal{H}=\mathrm{Sym}^{L}\left(\mathbb{C}^{N}\right)$) or $L$ fermions
($\mathcal{H}=\bigwedge^{L}\left(\mathbb{C}^{N}\right)$ ),
i.e.
\begin{eqnarray}
U.\ket{v_1}\vee\cdots\vee\ket{v_L}&=&U\ket{v_1}\vee\cdots\vee U\ket{v_L},\quad
\ket{v_1}\vee\cdots\vee\ket{v_L}\in\mathrm{Sym}^{L}\left(\mathbb{C}^{N}\right)
\nonumber \\
U.\ket{v_1}\wedge\cdots\wedge\ket{v_L}&=&U\ket{v_1}\wedge\cdots\wedge U\ket{v_L},
\quad \ket{v_1}\wedge\cdots\wedge\ket{v_L}\in\bigwedge^{L}\left(\mathbb{C}^{N}\right)
\nonumber
\end{eqnarray}
In the latter case we demand that $N\geq L$. The whole formalism concerning
the structure and the relevance of critical points of $\left\Vert
\mu\right\Vert ^{2}$ is valid also in this case. Only differences one should
keep in mind is that the momentum map
$\mu:\mathbb{P}\left(\mathcal{H}\right)\rightarrow i\mathfrak{k}^\ast$ and
the corresponding map $\mu^{\ast}$ are given by slightly different
expressions than for distinguishable particles \cite{HSK12}. The momentum map
is given (up to an irrelevant multiplicative constant) by
\begin{equation}
\mu\left(\left[v\right]\right)=\rho_{\left[v\right]}-\frac{1}{N}I_{N}\,,
\end{equation}
where $\rho_{\left[v\right]}$ is the reduced one-particle density matrix.Note that since the considered states are fully (anti)symmetric $\mu([v])$
does not depend upon the choice of the subsystem which is omitted in the
partial trace. Just like for distinguishable particles we have the
map $\Psi:\mathbb{P}\left(\mathcal{H}\right)\rightarrow i\mathfrak{t}^{+}$ given
by:
\begin{equation}
\Psi\left(\left[v\right]\right)=\tilde{\rho}_{\left[v\right]}-\frac{1}{N}I_{N}\,,
\end{equation}
where $\tilde{\rho}_{[v]}$ is the diagonalized reduced density matrix with
the ordered spectra $\tilde{\rho}_{[v]}=
\mathrm{diag}\left(\lambda_{1},\lambda_{2}\ldots,\lambda_{N}\right)$,
$\lambda_{1}\geq\lambda_{2}\geq\ldots\geq\lambda_{N}$.

In order to find critical sets of $\mathrm{Var}[v]$ (or equivalently of $\left\Vert \mu\right\Vert ^{2}$),
we consider the (unique) intersection of the adjoint orbit trough $\mu\left(\left[v\right]\right)$ with
the positive Weyl chamber $\mathfrak{it}^{+}$. On the practical level
this amounts to consideration of the diagonalized $\alpha\in\Psi\left(\mathbb{P}\left(\mathcal{H}\right)\right)$
and the corresponding operator $\alpha^{\ast}$,
\begin{equation}
\alpha^{\ast}=\alpha\otimes I_{N}\otimes\ldots\otimes I_{N}+\text{other permutations}\,,
\end{equation}
where in each of $L$ terms there are $L-1$ identity operators and
one operator $\alpha$. Next one has to check which eigenstates
$v\in\mathcal{H}$ of $\alpha^{\ast}$ indeed give $\alpha$ under
the momentum map: $\alpha=\mu\left([v]\right)$. The Morse index at the critical point:
\begin{itemize}
\item equals $0$ when $\left[v\right]$ is minimal ($\left[v\right]\in\mu^{-1}\left(0\right)$),
\item is the same as the Morse index of the Hessian of $\mu_{\mu\left(\left[v\right]\right)}\left(\left[w\right]\right)=\frac{\bk w{\mu^{\ast}\left(\left[v\right]\right)w}}{\bk ww}$
treated as a function of $\left[w\right]$ reduced to $T_{\left[v\right]}G.[v]{}^{\perp\omega}\subset T_{\left[v\right]}\mathbb{P}\left(\mathcal{H}\right)$,
where $G=K^{\mathbb{C}}=SL(N,\,\mathbb{C})$ acts by the diagonal
action on $\mathcal{H}=\mathrm{Sym}^{L}\left(\mathbb{C}^{N}\right)$
or $\mathcal{H}=\bigwedge^{L}\left(\mathbb{C}^{N}\right)$.
\end{itemize}

\subsection{Two bosons and two fermions}
The calculations presented in section \ref{bipartite} for bipartite states of
distinguishable particles have been greatly simplified by using the canonical
form of a state with respect to $K$ action, i.e. Schmidt decomposition. It
is, however, not so well known that there are similar standard forms for the
diagonal action of $K=SU(N)$ on
$\mathcal{H}=\mathrm{Sym}^{2}\left(\mathbb{C}^{N}\right)$ and
$\mathcal{H}=\bigwedge^{2}\left(\mathbb{C}^{N}\right)$. Up to our knowledge
the following theorem was proved for the first time in 1944.

\begin{theorem}\label{congruence}(Loo-Keng Hua \cite{Hua44})
Let $M$ be a complex $N\times N$ matrix\footnote{Notice that because $M$ is a complex matrix these statements are non-trivial.}.
\begin{enumerate}
\item If $M$ is symmetric ($M^{t}=M$) then there exist a unitary matrix
$U$ such that
\begin{gather}
UMU^{t}=\mathrm{diag}(a_{1},\, a_{2},\ldots,a_{n}),\label{eq:Schmidt_bosons}
\end{gather}

\item If $M$ is antisymmetric ($M^{t}=-M$) then there exist a unitary matrix
$U$ such that
\begin{gather}
UMU^{t}=\left(\begin{array}{cc}
0 & a_{1}\\
-a_{1} & 0
\end{array}\right)\dotplus\ldots\dotplus\left(\begin{array}{cc}
0 & a_{k}\\
-a_{k} & 0
\end{array}\right)\dotplus0\dotplus\ldots\dotplus0.\label{eq:Schmidt_fermions}
\end{gather}
\end{enumerate}
where $a_{i}$ are non-negative square roots of eigenvalues of $MM^{\dagger}$.
\end{theorem}
The statement of the theorem for or an arbitrary complex matrix $M$, which is known as Youla decomposition, can be found in \cite{Youla61}. It is worth to mention that theorem \ref{congruence} was also proved independently in \cite{Bozony}, \cite{Fermiony} and then using symplecto-geometric methods in \cite{HSK12}. As a direct consequence a normalized state of two $N$-state bosons or fermions can be written in the form:
\begin{itemize}
\item For bosons
\begin{equation}
v_B=a_{1}\left(\ket 1\otimes\ket 1\right)+a_{2}\left(\ket 2\otimes\ket 2\right)+\ldots+a_{N}\left(\ket N\otimes\ket N\right),\,\label{eq:bosons}
\end{equation}

\item For fermions
\begin{equation}
v_F=a_{1}\left(\ket 1\wedge\ket 2\right)+a_{2}\left(\ket 3\wedge\ket 4\right)+\ldots+
a_{\left\lfloor \frac{N}{2}\right\rfloor }\left(\ket{2\left\lfloor
\frac{N}{2}\right\rfloor -1}\wedge\ket{2\left\lfloor \frac{N}{2}\right\rfloor -1}\right)\,,
\label{eq:fermions}
\end{equation}
\end{itemize}
In both cases the reduced density matrix is a diagonal matrix
\begin{eqnarray}
\rho_{\left[v_B\right]}=\mathrm{diag}\{a_1^2,a_2^2,\ldots,a_N^2\}.\\\nonumber
\rho_{\left[v_F\right]}=\mathrm{diag}\{a_1^2,a_1^2,a_2^2,a_2^2\ldots,a_{\left\lfloor \frac{N}{2}\right\rfloor }^2,a_{\left\lfloor \frac{N}{2}\right\rfloor }^2,0,\ldots,0\}
\end{eqnarray}
Next, according to \eqref{eig}, in order to find critical points of $\left\Vert \mu\right\Vert ^{2}$
we look at solutions of the equation:
\begin{equation}\label{bosonfermions}
\mu^{\ast}\left(\left[v\right]\right)v=\lambda v,
\end{equation}
where:
\begin{equation}
\mu^{\ast}\left(\left[v\right]\right)=\left(\rho_{\left[v\right]}-
\frac{1}{N}I_{N}\right)\otimes I_{N}+I_{N}\otimes\left(\rho_{\left[v\right]}
-\frac{1}{N}I_{N}\right)\,.
\end{equation}
In order to avoid confusion we will continue with calculations separately for
bosons and fermions
\paragraph{Two bosons}
Equation (\ref{bosonfermions}) reads
\begin{equation}
a_{i}\left(|a_{i}|^{2}-\frac{1}{N}\right)=\lambda a_{i},\quad i=1,\ldots,N\,,
\end{equation}
for some real $\lambda$. Taking into account the requirement of ordering
of the spectra of $\mu([v])$ and the normalization we get $N$ families
of (nonequivalent) critical states:
\begin{equation}
v_{k}=\frac{1}{\sqrt{k}}\sum_{i=1}^{k}\ket i\otimes\ket i\,,\, k=1,\ldots,N\,.
\end{equation}
Notice that states $v_k$ are exactly the same as those corresponding to the
critical sets for bipartite distinguishable case. The bosonic character of
states $v_k$ is revealed by calculation of the Morse indices which we now
show to be different. To this we notice that:
\begin{gather}
\left(T_{[v_{k}]}G.[v_{k}]\right)^{\bot\omega}=\mathrm{Sym}^2\mathcal{H}_k\\
\mathcal{H}_k=\mathrm{Span}_{\mathbb{C}}\left\{ \ket{m}\otimes \ket{n},\:\, m,n\in\{k+1,\ldots,N\},\, m\neq n\right\},
\end{gather}
By simple calculations
$\dim_\mathbb{R}\left(\left(T_{[v_{k}]}G.[v_{k}]\right)^{\bot\omega}\right)
=\left(N-k\right)(N-k+1)$.
Analogously, like in Equation~\eqref{eq:index dist}, all directions from
$\left(T_{[v_{k}]}G.[v_{k}]\right)^{\bot\omega}$ correspond to the decrease
of $\left\Vert \mu\right\Vert ^{2}$ and therefore:
\begin{equation}
\mathrm{ind}([v_{k}])=\left(N-k\right)(N-k+1)\,.
\end{equation}
As promised the index of $[v_{k}]$ differs from the case of distinguishable
particles even though the form of the state
$[v_{k}]\in\mathbb{P}\left(\mathcal{H}\right)$ that encodes the critical
orbit of $K$ is the same in both cases.

\paragraph{Two fermions}

By repeating the reasoning from the previous paragraph we
get $\left\lfloor \frac{N}{2}\right\rfloor $ families of nonequivalent critical
sets parametrized by the states
\begin{equation}
v_{k}=\frac{1}{\sqrt{k}}\sum_{i=1}^{k}\ket{2i-1}\wedge\ket{2i}\,,\quad k=1,\ldots,\left\lfloor \frac{N}{2}\right\rfloor \,.
\end{equation}
Note that $\mu^{-1}(0)$ is empty when $N$ is not even. The index of a
critical point is computed in the same fashion as for two distinguishable
particles and two bosons. One checks that:
\begin{gather}
\left(T_{[v_{k}]}G.[v_{k}]\right)^{\bot\omega}= \mbox{Span}\left\{ \ket
m\wedge\ket n,\, i\left(\ket m\wedge\ket n\right):\,
m,n\in\{2k+1,\ldots,N\}\right\} .
\end{gather}
It follows that
$\dim\left(T_{[v_{k}]}G.[v_{k}]\right)^{\bot\omega}=\left(N-2k\right)\left(N-2k-1\right)$.
It turns out that each direction from
$\left(T_{[v_{k}]}G.[v_{k}]\right)^{\perp}$ corresponds to the decrease of
$\left\Vert \mu\right\Vert ^{2}$ and thus:
\begin{equation}
\mathrm{ind}([v_{k}])=\left(N-2k\right)(N-2k-1)\,.
\end{equation}

Note that due to the existence of the Hodge pairing,
 \[
\star:\bigwedge^k\left(\mathbb{C}^N\right)\rightarrow\bigwedge^{N-k}\left(\mathbb{C}^N\right)\ \text{,}
\]
one can relate the diagonal actions of $\mathrm{SL}(N,\mathbb{C})$ on $\mathbb{P}\left(\bigwedge^k\left(\mathbb{C}^N\right)\right)$ and  $\mathbb{P}\left(\bigwedge^{N-k}\left(\mathbb{C}^N\right)\right)$ respectively. The above isomorphism is not unique. By choosing a scalar product on $\mathbb{C}^N$ and fixing the orientation on $\mathbb{C}^N$  we
can make it unique:
\[
\star(\ket{i_1} \wedge \ket{i_2} \wedge \ldots \wedge \ket{i_k})=\ket{i_{k+1}} \wedge \ket{i_{k+2}} \wedge \ldots \wedge \ket{i_N}\ \text{,}
\]
where $\left\{i_1,i_2,\ldots,i_N\right\}$ is the even permutation of $\left\{1,2,\ldots,N\right\}$ and vectors $\left\{\ket{1},\ket 2,\ldots, \ket N \right\}$ form the orthonormal basis of $\mathbb{C}^N$. Physically this procedure amounts to consideration of particle - hole duality  \cite{Pauli principle} and allows to transfer the results obtained for the case two $N$ - state fermions to the case of $N-2$, $N$ - state fermions.

\subsection{Arbitrary number of two-state bosons.}\label{two-state bosons}
\color{black} For $L$ two-state bosons or, equivalently, symmetric states of
$L$ qubits we have $\mathcal{H}=\mathrm{Sym}^{L}\left(\mathbb{C}^{2}\right)$
with $K=SU\left(2\right)$ and $G=SL\left(2,\,\mathbb{C}\right)$. We choose an
orthonormal basis of $\mathbb{C}^{2}$ spanned by $\ket 0$, $\ket 1$ and call
$\ket{1}$ excited state.
An arbitrary state $v\in\mathrm{Sym}^{L}\left(\mathbb{C}^{2}\right)$ can be
written as a linear combination of $L+1$ basis elements called the Dicke
states.
\begin{gather}
v=\sum_{k=0}^{L}a_{k}\ket{k,\, L}\,
\end{gather}
where $\sum_{k=0}^{L}|a_{k}|^{2}=1$. Recall that Dicke state $\ket{k,\,L}$ of $L$ bosons is a normalized symmetric state for which $k$ out of $L$ bosons are in the excited state. The corresponding reduced density matrix
can be explicitly computed in the basis $\left\{ \ket 1,\,\ket 0\right\} $:
\begin{equation}
\rho_{\left[v\right]}=\frac{1}{L}\left[\begin{array}{cc}
\sum_{k=0}^{L}k|a_{k}|^{2} & \sum_{k=0}^{L-1}\frac{\bar{a}_{k}a_{k+1}}{\sqrt{\binom{L}{k}\binom{L}{k+1}}}\\
\sum_{k=0}^{L-1}\frac{a_{k}\bar{a}_{k+1}}{\sqrt{\binom{L}{k}\binom{L}{k+1}}} & \sum_{k=0}^{L}\left(L-k\right)|a_{k}|^{2}
\end{array}\right]\,.
\end{equation}
In order to find $\mu^{-1}\left(0\right)$ (that is minimal critical points of
$\left\Vert \mu\right\Vert ^{2}$) one has to impose condition
$\rho_{\left[v\right]}=\frac{1}{2}\mathbb{I}$, which amounts to solving
equations
\begin{gather}
\sum_{k=0}^{L}k|a_{k}|^{2}=\frac{L}{2}\,,\\
\sum_{k=0}^{L-1}\frac{a_{k}\bar{a}_{k+1}}{\sqrt{\binom{L}{k}\binom{L}{k+1}}}=0\,.
\end{gather}
We found it a hard problem to solve these equations for the general situation
(except for the trivial case $L=2$ and $L=3$ -- see below). It is
much easier to consider non-minimal critical points of $\left\Vert \mu\right\Vert ^{2}$.
To this end it is sufficient to consider solution to the problem
\begin{equation}
\alpha^{\ast}v=\lambda v\,,
\end{equation}
where $\alpha^{\ast}$ and $\alpha=\left[\begin{array}{cc}
\lambda & 0\\
0 & -\lambda
\end{array}\right]$ with $\lambda\in(0,\frac{1}{2}]$. One easily computes that
\begin{equation}
\alpha^{\ast}\ket{k,\, L}=\lambda_{k}\ket{k,\, L}\,,\quad k=0,1,\ldots,L,
\end{equation}
where $\lambda_k=L-2k$. For the considered values of $\lambda$ all eigenvalues $\lambda_{k}$ are non-degenerate and therefore the
only candidates for critical states are $v_{k}$. One checks that
\begin{equation}
\rho_{[v_{k}]}=\left[\begin{array}{cc}
\frac{L-2k}{2L} & 0\\
0 & \frac{2k-L}{2L}
\end{array}\right].
\end{equation}
Clearly, for $k=0 \ldots \lfloor \frac{L}{2}\rfloor$, the mapping
$\rho_{[v_{k}]}$ has the desired property. Therefore we have the following
states parameterizing inequivalent families of SLOCC classes
\begin{equation}
v_{k}=\ket{k,\, L}\,,\, k= 0 \ldots \lfloor \frac{L}{2}\rfloor\,.
\end{equation}
In the following we assume that $k$ is from the above range. The
infinitesimal action of $G$ on $v_k$ can only add one additional excitations
and hence
\begin{equation}
\left(T_{[v_{k}]}G.[v_{k}]\right)^{\bot\omega}=\mbox{Span}\left\{ \ket{m,\, L},\, i\ket{m,\, L}\,,\quad m\in\{0,1,\ldots,k-2\}\cup\left\{ k+2,\ldots,L\right\} \right\} \,,
\end{equation}
where it is assumed that $k\geq2$. For states $v_{0}$
and $v_{1}$ we have
\begin{equation}
\left(T_{[v_{k}]}G.[v_{k}]\right)^{\bot\omega}=\mbox{Span}\left\{ \ket{m,\, L},\, i\ket{m,\, L}\,,\quad m\in\{k+2,\ldots,L\}\right\} \,,
\end{equation}
Remarkably not all directions from
$\left(T_{[v_{k}]}G.[v_{k}]\right)^{\bot\omega}$ cause the decrease of
$\left\Vert \mu\right\Vert ^{2}$. Closer look at the Hessian of function
$\mu_{\mu\left(\left[v\right]\right)}\left(\left[w\right]\right)=\frac{\bk
w{\mu^{\ast}\left(\left[v\right]\right)w}}{\bk ww}$ at arbitrary
$[w]=[v_{k}]$ restricted to $\left(T_{[v_{k}]}G.[v_{k}]\right)^{\bot\omega}$
reveals that $\left\Vert \mu\right\Vert ^{2}$ decreases in the directions
given by vectors $\ket{m,\, L}$ and $\ket{m,\, L}$ for $m\in\left\{ \lfloor
\frac{L}{2}+1\rfloor,\ldots, L\right\} $. Notice that since we are interested
only in $k\in\{ 0,\ldots, \lfloor \frac{L}{2}\rfloor\}$ we finally obtain
\begin{equation}
\mathrm{ind}([v_{k}])=2\left\lceil \frac{L}{2}\right\rceil.
\end{equation}

\section{Summary}

A geometric description of entanglement, or more generally, quantum
correlations presented in our paper proved to be very fruitful in analysis of
various aspects of composite systems. The main achievement is a universal
algorithm allowing for determination of inequivalent SLOCC families of pure
states. For qubit systems the algorithm is effective - we showed how it works
for two-, three-, and four-qubit systems. Since the canonical form of a state
is known for arbitrary number of qubits, similar calculations can be, in
principle, performed for such systems, although definitely with more effort. In
any case the algorithm can be implemented numerically.

The algorithm distinguishes in a natural way two types of SLOCC classes:
those which are parameterized by nontrivial closed $G$-orbits where
$G$ is the group of SLOCC transformations, and those which belong
to the null cone $N$, i.e. contain in their closures $0$ as a unique
closed $G$-orbit. The orbits of the first type exhaust almost all
$G$-orbits. From this point of view orbits in the null cone seem
to be not particularly interesting, but in fact it is not so. Indeed,
starting from the equivalence classes of pure states one can construct
a classification of mixed quantum states. Different classes of mixed
states are then defined as convex sums of projectors on pure states
from a particular class. Pure states classes of measure zero may give
rise on the level of mixed to sets of non-zero measure \cite{acin01}.

The main tool used in our approach was the momentum map and its geometric
properties - in particular convexity of its image (or its appropriate
restriction). The momentum map is a natural construction whenever
a group acts on a symplectic manifold in a symplectic manner. Such
a situation is characteristic not only for the case of quantum entanglement
for distinguishable particles where the appropriate group consist
of all local unitary transformations acting the space of multiparticle
states, but \textit{mutatis mutandis} also in the bosonic and fermionic
cases, when additional symmetry conditions are imposed upon states.
The geometric and group-theoretic origins of our algorithm make it
versatile enough to cover also these cases.

In the course of argumentation we exhibited also some additional interesting
properties of the momentum map useful in characterizing entanglement
of states by appropriately defined distance of the SLOCC orbits from
the origin and by counting directions in which entanglement can be
increased under non-SLOCC operations. Finally we showed its connection
with potentially measurable physical observables.

Recently, after completion of our work, we have been informed that Walter,
Doran, Gross, and Christandl \cite{Walter} are independently studying the
SLOCC Kirwan polytopes, which they call entanglement polytopes in the context
of classification of SLOCC classes of states. Their work uses similar
methods, however, we would like to note that there are also some conceptual
differences between both approaches.

\section{Acknowledgments}
We would like to thank Alan Huckleberry and Peter Heinzner for fruitful
discussions. We gratefully acknowledge the support of SFB/TR12 Symmetries and
Universality in Mesoscopic Systems program of the Deutsche
Forschungsgemeischaft, ERC Grant QOLAPS, a grant of the Polish National
Science Center under the contract number DEC-2011/01/M/ST2/00379 and Polish
MNiSW grant no. IP2011048471.


\begin{thebibliography}{10}

\bibitem{ChuangNielsen} M.A Nielsen and I.L Chuang, Quantum Computation and Quantum Information, Cambridge Univ. Press, Cambridge 2000.

\bibitem{Horodeccy} R. Horodecki et. al., Quantum entanglement, Rev. Mod. Phys. 81, 865–942 (2009).

\bibitem{Sudbery} A. Sudbery, On local invariants of pure three-qubit states, \textit{J. Phys. A} 34, 643 (2001).

\bibitem{Vrana} P. Vrana, Local unitary invariants for multipartite quantum systems,  \textit{J. Phys. A} 44, 115302 (2011).

\bibitem{sawicki11} A.~Sawicki, A.~Huckleberry, and M.~Ku\'{s}, Symplectic
    geometry of entanglement, \textit{Comm. Math. Phys.} 305, 441--468 (2011).

\bibitem{Vidal} G. Vidal, Entanglement monotones,\textit{ Journal of Modern Optics} 47, (1999).


\bibitem{M03}S.~Mukai, An Introduction to Invariants and Moduli, Cambridge
    Studies in Advanced Mathematics (No. 81) 2003.



\bibitem{klyachko07} A.~Klyachko, Dynamic symmetry approach to entanglement,
    \emph{Proceedings of the NATO Advanced Study Institute on Physics and Theoretical
    Computer Science}, IOS Press, Amsterdam, 2007.

\bibitem{Kirwan82}F. C. Kirwan, Cohomology of Quotients in Symplectic and
    Algebraic Geometry, Mathematical Notes, Vol. 31, Princeton Univ. Press,
    Princeton, 1984.


\bibitem{Ness84}L. Ness, A stratification of the null cone via the moment map
    {[}with an appendix by D. Mumford{]}, \textit{Amer. J. Math.} 106(6), 1281--1329
    (1984).

\bibitem{GS84}V. Guillemin and S. Sternberg, Convexity properties of the moment
    mapping, II, \textit{Invent. Math.} 77, 533--546 (1984).
		
		\bibitem{GS90}V. Guillemin and S. Sternberg, Symplectic Techniques in
    Physics, 2nd ed., Cambridge Univ. Press, Cambridge, 1990.
		
		\bibitem{hall03} B. C. Hall, Lie groups, {L}ie algebras, and
    representations, an elementary introduction, Springer, New York, 2003.
		
\bibitem{Atiyah82}Atiyah, M. F.: Convexity and commuting Hamiltonians, Bull.
    London Math. Soc. 14, 1--15 (1982).
		


\bibitem{GS82}V. Guillemin and S. Sternberg, Convexity properties of the moment
    mapping, \textit{Invent. Math.} 67, 491--513 (1982).
		
		\bibitem{brion87} M.~Brion, Sur l'image de l'application moment, \textit{Lect. Notes
    Math.} \textbf{1296}, 177--192 (1987).

\bibitem{GuliSjamaar06}V. Guillemin and R. Sjamaar, Convexity theorems for
    varieties invariant under a Borel subgroup, \textit{Pure Appl. Math.} Q. 2 (2006),
    no. 3, 637--653.
		
			\bibitem{BR80} A. Barut and B. R\c{a}czka, Theory of group
    representations and
    applications, PWN, Warszawa, 1980.
		
			\bibitem{oszmaniec12} M. Oszmaniec and M. Ku\'{s}, On detection of
    quasiclassical states, \textit{J. Phys. A} 45, 244034 (2012).
		
		
\bibitem{KN82}G. Kempf, L. Ness, The lengths of vectors in representation
    spaces, \textit{Lectures Notes in Mathematics}, vol. 732, 233--243 (1982).
		
		\bibitem{Milne05}J. S. Milne, Algebraic geometry, Taiaroa Publishing Erehwon, Version 5.00 (2005).
		
		\bibitem{Mumford77}D. Mumford, Stability of projective varieties, L'Enseignement Mathématique. Revue Internationale. IIe Série 23 (1): 39–110 (1977).
		
		
		
\bibitem{HH96}P. Heinzner and A. Huckleberry, K\"ahlerian potentials and
    convexity properties of the moment map, \textit{Invent. Math.} 126, 65--84 (1996).
		
		

		
		\bibitem{Higuchi03}A. Higuchi, A. Sudbery, and J. Szulc, One-qubit reduced
    states of a pure many-qubit state: polygon inequalities, \textit{Phys. Rev. Lett.}
    90, 107902 (2003).
		
		\bibitem{VDM03} F. Verstraete, J. Dehaene, and B. De Moor, Normal forms and
    entanglement measures for multipartite quantum states, \textit{Phys. Rev.} A 68,
    012103 (2003).

    \bibitem{VDDV02} F. Verstraete, J. Dehaene, B. De Moor, and H. Verschelde, Four
    qubits can be entangled in nine different ways, \textit{Phys. Rev.} A 65, 052112
    (2002).
		
		
		
\bibitem{SK11} A. Sawicki, M. Ku\'{s}, Geometry of the local equivalence of
    states, \textit{J. Phys. A} 44, 495301 (2011).
		
				\bibitem{DWC00}W. D\"{u}r, G. Vidal, and J. I. Cirac, Three qubits can be entangled
    in two inequivalent ways,  \textit{Phys. Rev. A} 62, 062314 (2000).
		
				\bibitem{SWK12}A. Sawicki, M. Walter, M. Ku\'{s}, When is a pure state of three
    qubits determined by its single-particle reduced density matrices? to be
    published (2012).
		
		
		\bibitem{HSK12} A. Huckleberry, A. Sawicki, and M. Ku\'{s}, Bipartite
    entanglement, spherical actions and geometry of local unitary orbits, to be
    published (2012).
		
		\bibitem{Hua44}Loo-Keng Hua. On the Theory of Automorphic Functions
of a Matrix Variable I-Geometrical Basis. \textit{American Journal of Mathematics},
Vol. 66, No. 3 (1944), pp. 470-488.

\bibitem{Youla61}D. C. Youla. A normal form for a matrix under the
unitary congruence group Canad. \textit{J. Math.} 13 (1961), 694-704.

\bibitem{Bozony} K. Eckert, J. Schliemann, D. Bru\ss, and M. Lewenstein, Quantum Correlations in. Systems of Indistinguishable Particles, \textit{Ann.
    Phys.} 299, 88--127 (2002).

\bibitem{Fermiony} J. Schliemann, J. I. Cirac, M. Ku\'{s}, M. Lewenstein, and
    D. Loss, Quantum Correlations in Two-Fermion Systems, \textit{Phys. Rev. A} 64, 022303 (2001).
		
\bibitem{Pauli principle} M. Altunbulak and A. Klyachko, The Pauli principle
    revisited, \textit{Comm. Math. Phys.} 282, 287--322 (2008).

\bibitem{acin01} A. Acin, D. Bru\ss, M. Lewenstein, and A. Sanpera,
    Classification of Mixed Three-Qubit States, \textit{Phys. Rev. Lett}. 87, 040401
    (2001).

\bibitem{Walter}M. Walter, B. Doran, D. Gross, M. Christandl, Entanglement Polytopes, arXiv:1208.0365, (2012).









\bibitem{Schmidt08}A. H. W. Schmitt, Geometric Invariant Theory and Decorated Principal Bundles, Zurich Lectures in Advanced Mathematics, ISBN 978-3-03719-065-4, (2008).





























\end{thebibliography}
\end{document}